\documentclass{aa}  
\usepackage[english]{babel}
\usepackage{amsmath}
\usepackage{graphicx}
\usepackage{natbib}
\usepackage[colorlinks=true, allcolors=blue]{hyperref}
\usepackage{txfonts}
\begin{document}

 \title{HALO 
 }
 \subtitle{I: Photometric continuum reverberation mapping of Fairall~9}
 \titlerunning{UV-optical lag-spectrum of Fairall~9 from disk reverberation mapping}

   \author{Amit Kumar Mandal \inst{1,\thanks{amitastro.am@gmail.com}}
    \and
    Francisco Pozo Nu\~nez \inst{2,\thanks{francisco.pozon@gmail.com}}
          \and
   Vikram Kumar Jaiswal \inst{1} 
   \and 
   Mohammad Hassan Naddaf\inst{3}
   \and 
    Bo\.zena Czerny \inst{1,\thanks{bcz@cft.edu.pl}}
          \and
          Swayamtrupta Panda\inst{4,\thanks{Gemini Science Fellow}}
          \and
Paulina Karczmarek\inst{5}
          \and
Grzegorz Pietrzy\' nski\inst{5,6}
\and
Shivangi Pandey\inst{7,15}
\and
B. M. Peterson\inst{8, \thanks{Retired.}}
\and
{Michal Zaja\v{c}ek}\inst{9}
\and
{Michal Dov\v{c}iak}\inst{10}
\and
{Vladimir Karas}\inst{10}
\and
{Weronika Narloch}\inst{5}
\and
{Miros\l{}aw Kicia}\inst{5}
\and
{Marek G\'orski}\inst{5}
\and
{Miko\l{}aj Ka\l{}uszy\'nski}\inst{5}
\and
{Gergely Hajdu}\inst{5}
\and
{Piotr Wielg\'orski}\inst{5}
\and
{Bart\l{}omiej Zgirski}\inst{6}
\and
{Cezary Ga\l{}an}\inst{5}
\and
{Wojciech Pych}\inst{5}
\and
{Rados\l{}aw Smolec}\inst{5}
\and
{Karolina B\k{a}kowska}\inst{11}
\and
{Wolfgang Gieren}\inst{6,12}
\and
{Pierre Kervella}\inst{13,14}
  }

   \institute{Center for Theoretical Physics, Polish Academy of Sciences, Al. Lotnik\'ow 32/46, 02-668 Warsaw, Poland
%              \email{}
\and
Astroinformatics, Heidelberg Institute for Theoretical Studies, Schloss-Wolfsbrunnenweg 35, 69118 Heidelberg, Germany
    \and 
    Institut d'Astrophysique et de Géophysique, Université de Liège Allée du six août 19c, B-4000 Liège (Sart-Tilman), Belgium
    \and
    International Gemini Observatory/NSF NOIRLab, Casilla 603, La Serena, Chile
    \and
    Nicolaus Copernicus Astronomical Center, Polish Academy of Sciences, Bartycka 18, 00-716 Warszawa, Poland
    \and
    Universidad de Concepci\'on, Departamento de Astronom\'ia, Casilla 160 \textendash C, Concepci\'on, Chile
    \and
    Aryabhatta Research Institute of Observational Sciences, Nainital\textendash263001, Uttarakhand, India
    \and
    c/o Tracy L. Turner, 205 South Prospect Street, Granville, OH 43023, USA
    \and
    Department of Theoretical Physics and Astrophysics, Faculty of Science, Masaryk University, Kotlá\v{r}ská 2, 611 37 Brno, Czech Republic
    \and
    Astronomical Institute of the Czech Academy of Sciences, Bo\v{c}n\'i II 1401, CZ-14100 Prague, Czech Republic
    \and
    Institute of Astronomy, Faculty of Physics, Astronomy and Informatics, Nicolaus Copernicus University, ul. Grudzi\k{a}dzka 5, 87-100 Toru\'n, Poland
    \and
    Millenium Institute of Astrophysics, Avenue Libertador Bernardo O’Higgins 340, Casa Central, Santiago, Chile
    \and
    LIRA, Observatoire de Paris, Université PSL, Sorbonne Université, Université Paris Cité, CY Cergy Paris Université, CNRS, 92190 Meudon, France
    \and
    French-Chilean Laboratory for Astronomy, IRL 3386, CNRS and U. de Chile, Casilla 36-D, Santiago, Chile
    \and
    Department of Applied Physics/Physics, Mahatma Jyotiba Phule Rohilkhand University, Bareilly\textendash243006, India
    }

   \date{Received XXX; accepted XXX}

  \abstract
   {We investigate the origin of inter-band continuum time delays in active galactic nuclei (AGNs) to study the structure and properties of their accretion disks.}
   {We aim to measure the inter-band continuum time delays through photometric monitoring of Seyfert galaxy Fairall~9 to construct the lag-spectrum.
    Additionally, we explain the observed features in the Fairall~9 lag-spectrum and discuss the potential drivers behind them, based on our newly collected data from the Obserwatorium Cerro Murphy (OCM) telescope.}
   {We initiated a long-term, continuous AGN photometric monitoring program in 2024, titled 'Hubble constant constraints through AGN Light curve Observations' (HALO) using intermediate and broad band filters. Here, we present the first results from HALO, focusing on photometric light curves and continuum time-delay measurements for Fairall~9. To complement these observations and extend the wavelength coverage of the lag-spectrum, we also reanalyzed archival Swift light curves and spectroscopic data available in the literature.}
   {Using HALO and Swift light curves, we measured inter-band continuum delays to construct the lag-spectrum of Fairall~9. Excess lags appear in the $u$ and $U$ bands (Balmer continuum contamination) and in the $I$ band (Paschen jump/dust emission from the torus). Overall, the lag-spectrum deviates significantly from standard disk model predictions.}
  {We find that inter-band delays deviate from the power-law, $\tau_{\lambda} \propto \lambda^{\beta}$ due to BLR scattering, reprocessing, and dust contributions at longer wavelengths. Power-law fits are therefore not well suited for characterizing the nature of the time delays.}

\maketitle

\section{Introduction}

Active galactic nuclei (AGNs) are among the most luminous objects in the Universe, where matter accreting onto the central supermassive black hole (SMBH) forms an optically thick and geometrically thin standard accretion disk \citep[SSD;][]{SS1973}. According to the lamp-post model \citep{1989ESASP.296..945G, Matt1991}, this disk is illuminated by a compact ionizing X-ray source. The intercepted radiation is subsequently reprocessed into UV–optical emission, with progressively larger disk radii contributing to longer wavelengths as the disk temperature decreases outward, following the relation $T \, \sim \, R_{\rm disk}^{-3/4}$ \citep{collier1999}.

To probe the structure of such accretion disks, several studies have employed photometric continuum reverberation mapping \citep[continuum-RM;][]{2005ApJ...622..129S, 2014MNRAS.444.1469M, 2014ApJ...788...48S, 2017ApJ...836..186J, 2018ApJ...862..123M, 2019ApJ...880..126H, 2020ApJS..246...16Y, 2020ApJ...903..112D, 2021ApJ...922..151K, 2022MNRAS.511.3005J, 2022ApJ...940...20G, 2023ApJ...947...62K, 2024ApJ...961...93S, edelson2024, 2025ApJ...985...30M, 2025MNRAS.541..642P, 2025A&A...700L...8P}. These investigations have revealed several key outcomes. First, inter-band time delays are broadly found to follow a power-law dependence on wavelength, $\tau_{\lambda} \, \propto \, \lambda^{\beta}$, with a slope of $\beta \sim 4/3$ \citep{2021ApJ...922..151K, 2022ApJ...940...20G, 2023ApJ...947...62K, 2025ApJ...985...30M}, in good agreement with the predictions of the SSD model. This consistency now extends even to the highest redshift probed through continuum-RM ($z=2.7$; \citealt{2025A&A...700L...8P}). Nevertheless, a number of continuum-RM studies have reported accretion-disk sizes and wavelength-delay slopes that deviate from the SSD prediction \citep[e.g.,][]{2016ApJ...821...56F, 2017ApJ...835...65S, lawther2018, 2023A&A...672A.132F, 2023MNRAS.519.3366M}, suggesting that the structure and reprocessing properties of some AGN disks may be more complex than the simple model assumes.

Second, the continuum-emitting region size inferred from SSD-based power-law fits is consistently measured to be about three to five times larger than predicted, a discrepancy often referred to as the 'disk-size anomaly' \citep{2016ApJ...821...56F, 2023ApJ...947...62K, 2025ApJ...985...30M}. This mismatch is now thought to arise from contamination due to broad line region (BLR) scattering and reprocessing, which can increase the UV–optical time delays \citep{Korista_Goad_2001, netzer2022}. Alternatively, as discussed by \citet{2021ApJ...907...20K, kammoun2021, kammoun2023}, an enlarged disk size at a given wavelength can arise from the significant vertical extent of the X-ray corona. Interestingly, these continuum-emitting sizes also show correlations with AGN luminosity and BLR sizes, suggesting that continuum time delays measured from photometric light curves may provide a promising avenue for black hole mass measurements once established over a large dynamic range in luminosity \citep{2023ApJ...948L..23W, 2024ApJ...968L..16P, 2025ApJ...985...30M}.

Third, observed lag-spectra reveal a significant excess lag around the 3646 {\AA} Balmer jump \citep{cackett2018}, with an additional potential excess around the 8206 {\AA} Paschen jump \citep{GuoH_etal_2022}. These excess lags strongly indicate contamination from the BLR, causing the inter-band time delays to deviate significantly from a simple power-law wavelength dependence. 

Recently, we demonstrated that photometric continuum time delays provide a powerful method for estimating the Hubble constant ($H_0$) by simultaneously fitting the lag-spectrum and the spectral energy distribution (SED) of AGNs \citep[see][for details]{2025A&A...702A..92J}. Our study, based on the well-studied AGN, NGC~5548, yielded an $H_0$ estimate of $66.9^{+10.6}_{-2.1}$ km s$^{-1}$ Mpc$^{-1}$ with an uncertainty of about 10–15\%. Although this error is relatively large derived from a single object, the approach shows great promise, i.e., by extending such measurements to a large sample of AGNs spanning a wide range of redshifts, the uncertainty can be significantly reduced, thereby offering a new avenue to address the ongoing Hubble tension \citep{eleonora_2022}.

Motivated by this potential, we have initiated a long-term AGN photometric monitoring program at the Obserwatorium Cerro Murphy (OCM) 60 cm telescope in Chile, namely 'Hubble constant constraint through AGN Light curves Observation' (HALO), which began in 2024 and employs multi-filter photometry using the Str\"omgren $u$, $v$, $b$, $y$ and Johnson–Cousins $I$ bands to probe continuum time delays. As part of HALO, we have completed observations of one target, the bright Seyfert galaxy Fairall~9.

Fairall~9 is a well-known and extensively studied AGN. It was first identified during a survey for galaxies with bright, compact nuclei in the southern sky \citep{fairall1977}, and soon attracted attention because of its relative brightness and proximity, with a redshift of $z=0.046145$. Over the years, multi-wavelength observations spanning the X-ray, ultraviolet (UV), optical, and infrared (IR) bands have provided a comprehensive characterization of its physical properties. In particular, X-ray spectral studies have shown that Fairall~9 is relatively simple in structure, often classified as a 'bare AGN' since it lacks the strong warm absorbers commonly detected in many other Seyfert galaxies \citep{patrick2011,emmanopoulos2011}. Its intrinsic variability has further established it as a valuable target for long-term monitoring, motivating extensive campaigns that have been conducted over several decades \citep[e.g.,][]{rodriguez1997,1998PASP..110..660P}. Most notably, \citet{edelson2024} reported the results of an exceptionally dense, 1.8-year daily monitoring program with Swift, while \citet{hernandez2020} analyzed the first year of the same campaign, supplemented by ground-based observations. These studies provide an invaluable reference for our work, enabling a comparison between space-based and ground-based monitoring approaches.

In this paper, we present our optical photometric monitoring of Fairall~9 using 4 Str\"omgren and 1 Johnson-Cousins filters. Our current analysis focuses on constructing the lag-spectrum of this source, which will form the basis for subsequent estimation of $H_0$ within the broader HALO framework. The paper is organized as follows: Section~\ref{ss:obs} outlines the observations and data reduction procedures. Section~\ref{ss:analysis} details the flux variability, host-galaxy flux estimation, and time-series analysis. The results are presented in Section~\ref{res}, followed by a discussion in Section~\ref{ss:dis}. Finally, Section~\ref{ss:summ} provides a summary of the main findings.

\begin{figure}
    \centering
   \includegraphics[width=1\linewidth]{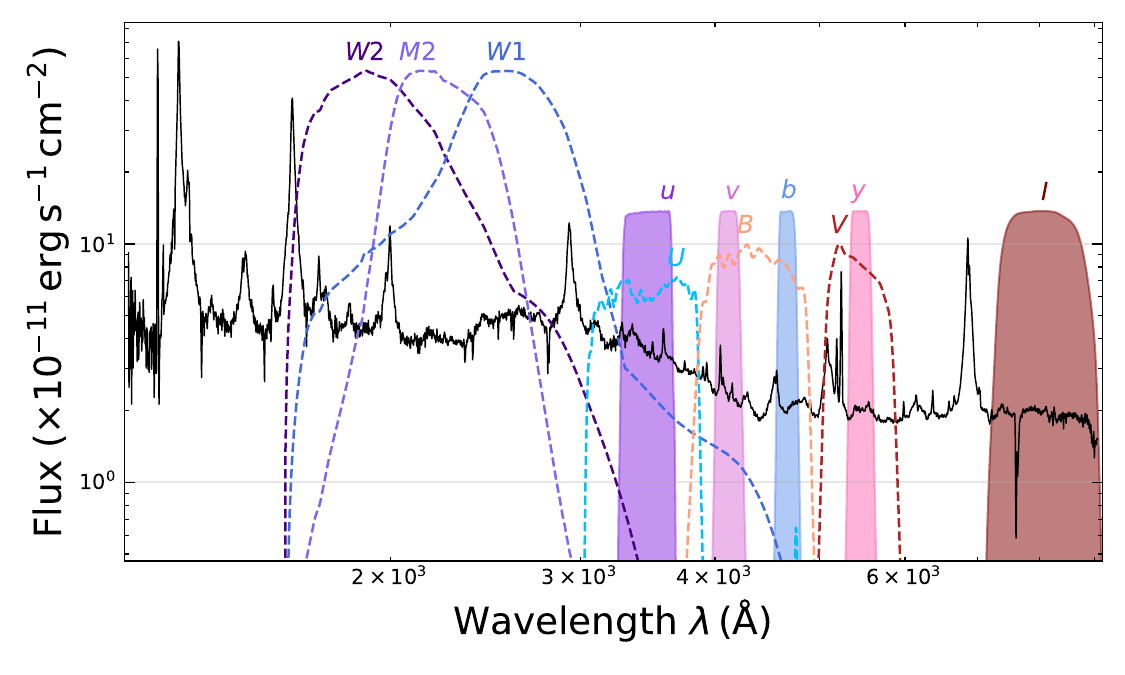}
    \caption{UV and optical spectra of Fairall~9 obtained with the HST Faint Object Spectrograph on 1993 January 21 and with the ESO 1.5 m telescope on 1994 July 14, respectively. Overplotted are the transmission curves of the OCM filters ($u$, $v$, $b$, $y$, and $I$) used for photometric observations in our HALO program, along with the transmission curves of Swift-UVOT filters ($W2$, $M2$, $W1$, $U$, $B$, and $V$) employed in constructing the photometric light curves in the literature.}
    \label{fig:spec}
\end{figure}

 \begin{figure*}
\sidecaption
  \includegraphics[width=12cm]{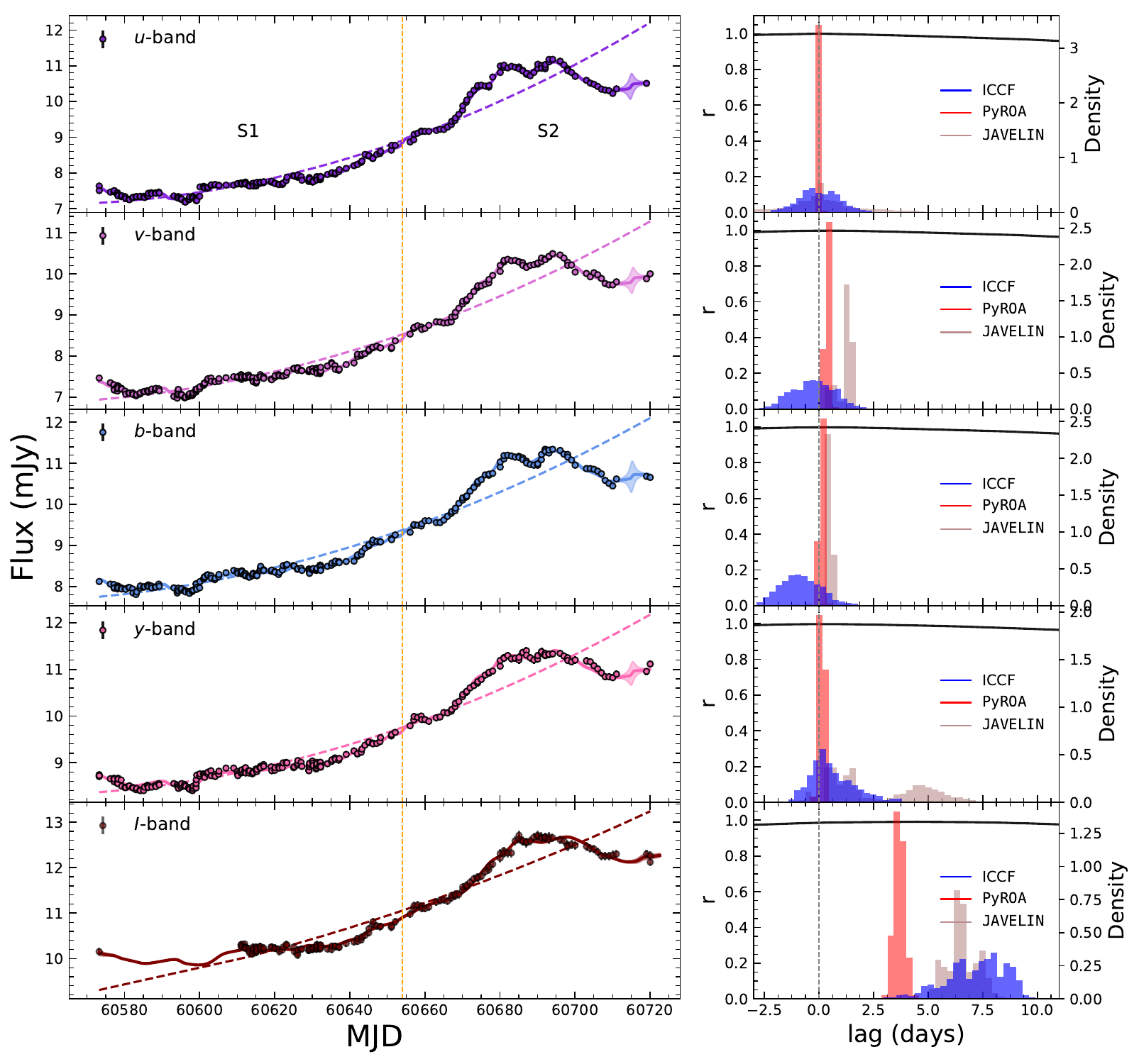}
     \caption{Light curves from OCM monitoring and lag analysis. Left: Light curves of Fairall~9 in the $u$, $v$, $b$, $y$, and $I$ filters, shown from top to bottom, respectively. In each panel, the dashed lines in various colors indicate second-order polynomial detrending, while the solid lines represent the best-fit light curves obtained from {\tt PyROA}. The full light curve is divided into two segments, S1 and S2, separated by an observational gap indicated by the vertical dashed orange line. Right: Distribution of the cross-correlation coefficient relative to the $u$ band (solid black line) derived using the ICCF method for the full, non-detrended original light curve. The blue histogram shows the cross-correlation centroid distribution from ICCF, the red histogram represents the lag probability distribution obtained from {\tt PyROA}, and the brown histogram shows the corresponding distribution from {\tt JAVELIN}. A vertical dashed gray line marks the reference point at $\tau$ = 0 days.}
     \label{fig:lc}
\end{figure*}

\section{Observations and Data reduction}
\label{ss:obs}

The core objective of HALO is to perform simultaneous modeling of inter-band continuum time delays and SEDs for a well-selected sample of AGNs across a broad luminosity and redshift range. Observations for this project are primarily conducted using meter-class robotic telescopes at the Rolf Chini Cerro Murphy Observatory ('OCM' from Polish: Obserwatorium Cerro Murphy), an international astrophysical project, hosted in European Southern Observatory (ESO) in Chile, and operated by the Nicolaus Copernicus Astronomical Center of the Polish Academy of Sciences in Warsaw, Poland. A detailed overview of the HALO project will be presented in Pozo Núñez et al. (in prep.).
Below, we describe the photometric monitoring campaign targeting Fairall~9.

\subsection{Observations}

The monitoring was conducted between September 20, 2024, and February 17, 2025, using the Str\"omgren $u$, $v$, $b$, and $y$ filters to trace the AGN's optical continuum in spectral regions minimally affected by emission lines. To extend coverage toward the red end of the optical spectrum, the Johnson-Cousins $I$ filter was also used. The main properties of Fairall~9 are summarized in Table~\ref{tab:prop}.

To ensure that our photometric measurements are free from contamination by broad emission lines, we carefully selected four Str\"omgren intermediate-band filters along with the Johnson-Cousins $I$ filter. This selection was made specifically to minimize contamination from any prominent broad emission features within the filter bandpasses. In contrast to many previous continuum$-$RM studies, which often lacked this level of precision in filter selection, our approach prioritizes minimizing contamination from the BLR while maximizing the contribution from the underlying accretion disk continuum. We illustrate the chosen filter transmission curves in Figure~\ref{fig:spec} overlaid on the optical spectrum of Fairall~9 \citep{edelson2024}. This approach provides an optimal framework to evaluate the consistency of our observations with predictions from the standard disk model and to explain the disk size anomalies reported in earlier studies.

\subsection{Data reduction}

For photometric data reduction, we followed standard procedures, including bias subtraction, dark current removal, flat-fielding, astrometric calibration, and distortion correction. These steps were carried out using a combination of standard {\tt IRAF}\footnote{{\tt IRAF} is distributed by the National Optical Astronomy Observatory, which is operated by the Association of Universities for Research in Astronomy (AURA) under cooperative agreement with the National Science Foundation.} \citep{1986SPIE..627..733T} tasks and custom-developed routines, following the detailed methodology described in \citet{2017PASP..129i4101P}.

The light curves were then extracted using aperture photometry. To optimize the signal-to-noise ratio (S/N) and minimize flux scatter, photometry was first performed on all stars in the field across a range of aperture sizes. From these measurements, a photometric curve of growth was constructed using the instrumental magnitudes at different apertures to evaluate the combined contributions from the host galaxy and the central engine of Fairall~9. Field stars were subsequently selected according to brightness, proximity to the AGN, and the standard deviation ($\sigma_*$) of their light curves.
 
Small apertures (1.0$-$2.0 arcsec) yielded higher scatter due to PSF sensitivity and incomplete flux inclusion, while large apertures (15$-$20 arcsec) increased $\sigma_*$ as a result of elevated background contamination. In general, $\sigma_*$ was found to be brightness-dependent. To ensure photometric stability, we selected reference stars with brightness levels similar to the AGN and located within 25 arcminutes of the target. From these, the top 20\% with the lowest $\sigma_*$ values were retained, yielding a final set of 12 reference stars. Subsequently, relative light curves were constructed by normalizing the AGN flux to the ensemble of selected reference stars.  The final AGN light curve corresponds to the average flux calibrated against this reference set.

The absolute flux calibration was performed using non-variable stars flagged for high quality in the DES Data Release~2 catalogue \citep{2021ApJS..255...20A}, located in the same field as Fairall~9. 
The calibration accounts for both atmospheric extinction \citep{2011A&A...527A..91P} and Galactic foreground extinction. The latter was derived from the \citet{2011ApJ...737..103S} recalibration of the dust maps by \citet{1998ApJ...500..525S}, using NED\footnote{\url{https://ned.ipac.caltech.edu/}} extinction values at the source coordinates with $E(B-V) = 0.023$. These values were interpolated to the effective wavelengths of our filters. The final light curve data in $u$, $v$, $b$, $y$, and $I$ bands are provided in Table~\ref{tab:lc}.

\begin{table*}[]
\centering

 \caption{Basic properties of Fairall~9}
 \label{tab:prop}

\resizebox{18cm}{!}{
\fontsize{8pt}{8pt}\selectfont
\begin{tabular}{ccccccccc} \hline \hline

  $z$ & log $L_{AGN,5100}$ & $F_{AGN,5100}$ (1+$z$) & $\tau_{\mathrm{cent, H\beta}}$  & $\sigma_{\mathrm{line,H\beta}}$ &  FWHM  & $M_{\text{BH}}$ & log$\lambda_{\text{Edd}}$ & $i$
  \\  
& $\mathrm{erg \, s^{-1}}$ & $\times 10^{-15} \, \mathrm{erg \, s^{-1} \, cm^{-2} \, {\AA}^{-1}}$ & days & $\mathrm{km \, s^{-1}}$ & $\mathrm{km \, s^{-1}}$ & $\times 10^8 M_{\odot}$ &  & degree  \\
(1) & (2) & (3) & (4) & (5) & (6) & (7) & (8) & (9)
\\ \hline \\
0.046145 & $43.92 \pm 0.05^{(a)}$ & $2.95 \pm 0.23^{(a)}$ & $17.4_{-4.3}^{+3.2 \, (b)}$ & $3787 \pm 197^{(b)}$ & $6901 \pm 707^{(b)}$ & $2.18 \pm 0.46$ & $-1.551 \pm 0.104$ & $35.0_{-5.6}^{+3.4 \, (c)}$ \\ \\

\hline
\end{tabular}
}
\tablefoot{Columns are: (1) redshift $z$, (2) AGN luminosity at 5100 {\AA}, (3) AGN monochromatic flux at 5100 {\AA}, (4) H$\beta$ BLR lag in rest-frame, (5) H$\beta$ line dispersion, (6) FWHM of  H$\beta$ line, (7) black hole mass measured from $\tau_{\mathrm{cent, H\beta}}$ and $\sigma_{\mathrm{line,H\beta}}$ with virial factor $f_{BLR}=4.47$ \citep{2015ApJ...801...38W}, (8) logarithm of Eddington ratio, and (9) inclination angle. References are: (a) \citet{bentz2013}, (b) \citet{peterson2004}, and (c) \citet{2012ApJ...758...67L}.}

\end{table*}

\begin{table}
\centering
\caption{Photometric light curves data from HALO}
\label{tab:lc}
\centering
\begin{tabular}{ccccc} \hline \hline

 MJD& $F$ & $F_{err}$ & filter \\
 & mJy & mJy &    \\
(1) & (2) & (3) & (4) 
\\ \hline

 60573.252254 & 7.6334 & 0.0413 & u \\
 60573.265551 & 7.5059 & 0.0406 & u \\
 60576.229650 & 7.4634 & 0.0404 & u \\

... & ... & ... & ...  \\
... & ... & ... & ...  \\

\hline

\end{tabular}
\vspace{0.01cm}

\tablefoot{Columns are: (1) Modified Julian Date, (2) flux, (3) error in flux, and (4) filter name. Full data table is available at the CDS. Corresponding light curves are shown in Figure~\ref{fig:lc}.}

\end{table}

\subsection{Spectrum data}

In addition to our photometric monitoring of Fairall~9, we also compiled spectroscopic data from the literature, as such data are essential for constructing the SED and assessing potential BLR contamination in the measured continuum inter-band delays. Specifically, we incorporated UV spectrum observed with the  \textit{Hubble Space Telescope} (HST) Faint Object Spectrograph obtained on 1993 January 21 \citep{edelson2024}. 
Furthermore,  an optical spectrum is available from the optical monitoring campaign using the European Southern Observatory (ESO) 1.5 m telescope equipped with the Boller \& Chivens spectrograph and a CCD detector \citep{1997ApJS..112..271S}, conducted in synergy with the eight-month monitoring campaign using the International Ultraviolet Explorer  \citep[IUE;][]{1997ApJS..110....9R}.

For our study, we utilized the combined UV and optical spectrum of Fairall~9 which was presented by \citet{edelson2024}. This spectrum is shown in Figure~\ref{fig:spec}.

\subsection{Additional photometric data from Swift}

We collected additional light curve data of Fairall~9 from the literature. For this purpose, we adopted the homogeneous and relatively uninterrupted set of Swift observations assembled by \citet{edelson2024}, which includes hard- and soft- X-ray light curves from the X-Ray Telescope (XRT) as well as UV-optical light curves in the $W2$, $M2$, $W1$, $U$, $B$, and $V$ bands from the Ultraviolet/Optical Telescope (UVOT). We did not include the dataset from \citet{hernandez2020}, as those Swift observations were already incorporated in the compilation of \citet{edelson2024}. Likewise, we followed \citet{edelson2024} in excluding data from the Las Cumbres Observatory Global Telescope  (LCOGT) network, and used Swift light curve data up to MJD $\sim$58900 in order to maintain the homogeneity and uniformity of the light curves.

\section{Analysis}
\label{ss:analysis}

\begin{table*}
\centering

 \caption{Light curve properties and results from the variability analysis}
 \label{tab:var}

\resizebox{18cm}{!}{
\fontsize{7pt}{7pt}\selectfont
\begin{tabular}{lccccccccc} \hline \hline

Filter & $\lambda_{\text{eff}}$ & N & $\Delta t$ &  $<F>$ & $\sigma$ & $F_{\mathrm{var}}$  & $R_{\mathrm{max}}$ & $F_{\text{AGN}}$  & $F_{\text{gal}}$  
\\ 
 & ({\AA}) & & (days) & (mJy) &  (mJy) & & & (mJy) & (mJy) \\
(1) & (2) & (3) & (4) & (5) & (6) & (7) & (8) & (9) & (10)
\\ \hline \\

$u$ & $3468\pm145$ & 236 & 0.25 & $8.478 \pm 0.046$   & 1.3085 & $0.1542\pm0.0004$  & 1.556 & $7.91 \pm 0.10$ & $0.571 \pm 0.026$  \\
$v$ & $4120\pm92$ & 240 & 0.23 &  $8.141 \pm 0.057$   & 1.1378 & $0.1396\pm0.0004$  & 1.504 & $6.96 \pm 0.13$ & $1.179 \pm 0.061$ \\
$b$  & $4668\pm75$ & 239 & 0.24 & $8.980 \pm 0.062$  &  1.1454 & $0.1273\pm0.0004$  & 1.452 & $7.00 \pm 0.17$ & $1.977 \pm 0.102$ \\
$y$  & $5460\pm90$ & 241 & 0.23 & $9.412 \pm 0.067$   & 0.9976 & $0.1057\pm0.0005$  & 1.357  & $6.13 \pm 0.31$ & $3.285 \pm 0.248$ \\
$I$  & $8105\pm465$ & 164 & 0.77 & $11.082 \pm 0.086$ & 0.9660 & $0.0868\pm0.0006$  & 1.262 & $6.75 \pm 0.22$ & $4.331 \pm 0.214$ \\
\hline

\end{tabular}
}

\tablefoot{Columns are: (1) name of filter, (2) effective wavelength of the filter with the rms width of the transmission curve representing the filter’s wavelength uncertainty, (3) number of data points in the light curve, (4) median cadence value of the light curve, (5) mean flux value, (6) standard deviation of light curve, (7) normalized excess variance, (8) maximum to minimum flux ratio, (9) median AGN flux, and (10)  median host-galaxy flux. The AGN and host-galaxy fluxes are measured from FVG method.
    }

\end{table*}

\begin{figure}
    \centering
   \includegraphics[width=0.9\linewidth]{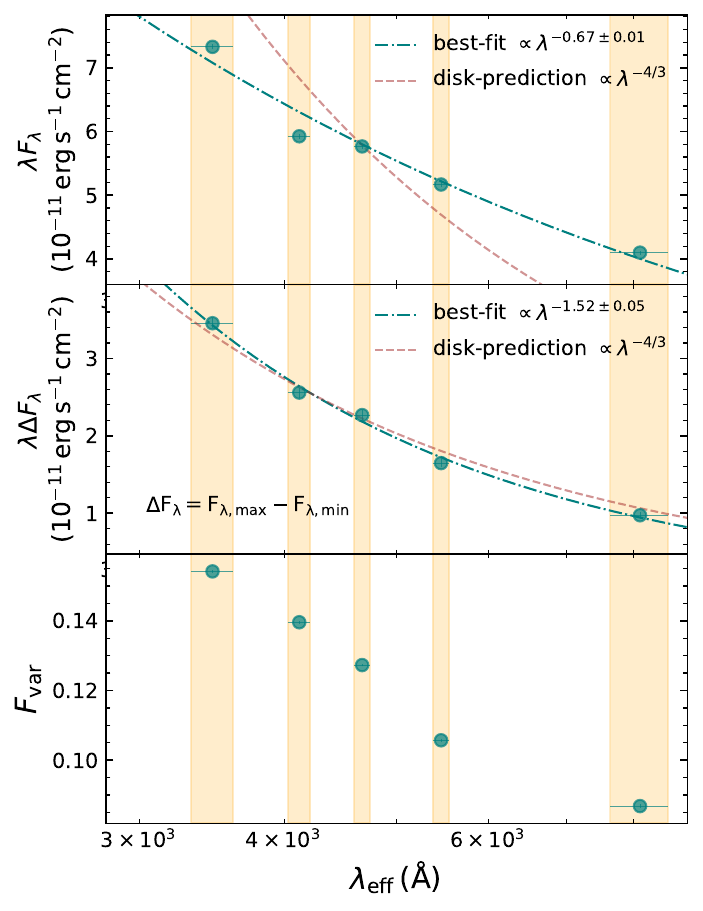}
    \caption{Results from photometric flux variability analysis. Top: Optical SED of Fairall~9 derived from the mean fluxes in the $u$, $v$, $b$, $y$, and $I$ filters. Middle: differential SED constructed from the difference between the maximum and minimum fluxes in the corresponding filters. Bottom: Normalized excess variance, $F_{\mathrm{var}}$ as a function of wavelength. In both the top and middle panels, the dot-dashed teal line represents the best-fit power law with a free spectral index, while the dashed brown line corresponds to a power law with the index fixed at $-4/3$, as predicted by the standard thin accretion disk model. The orange shaded regions indicate the $\pm 1 \sigma$ range, representing the wavelength uncertainty of each filter, as quantified by the rms width of its transmission curve.}
    \label{fig:fvar}
\end{figure}

\subsection{Flux variability}
We present the light curves of Fairall~9 in Figure \ref{fig:lc} in the $u$, $v$, $b$, $y$, and $I$ filters.  To quantify the flux variability of the source, we computed the normalized excess variance, $F_{\mathrm{var}}$ \citep{1997ApJS..110....9R, vaughan2003}, defined as 

\begin{equation}
F_{\mathrm{var}} = \sqrt{\frac{\sigma^2 - \bar{\epsilon}_{\mathrm{err}}^2}{<F>^2}},
\end{equation}
where $<F>$ is the mean flux over the light curve. The sample variance, $\sigma^2$, and the mean square measurement uncertainty, $\bar{\epsilon}^{2}_{\mathrm{err}}$, are calculated as  
\begin{equation}
\sigma^{2} = \frac{1}{N-1}\sum_{i=1}^{N}(F_{i} - <F>)^{2},
\end{equation}
\begin{equation}
\bar{\epsilon}^{2}_{\mathrm{err}} = \frac{1}{N}\sum_{i=1}^{N}{\epsilon^{2}_{i}},
\end{equation}
where $\epsilon_i$ denotes the uncertainty associated with each individual flux measurement. The uncertainties in $F_{\mathrm{var}}$ were estimated following the methodology described in \citet{2002ApJ...568..610E} and \citet{2021MNRAS.502.2140M}. Additionally, we calculated the maximum to minimum flux ratio, defined by the parameter $R_{\text{max}}$. The obtained variability properties are summarized in Table \ref{tab:var}.

In the top panel of Figure \ref{fig:fvar}, we present the SED in optical bands derived from the mean fluxes and their corresponding uncertainties, as obtained from the multi-band light curves. All flux measurements were corrected for Galactic extinction. A standard thin accretion disk model predicts an SED that follows the relation $\lambda F_{\lambda} \propto \lambda^{-4/3}$. To test this, we first fit the observed SED with a power-law function, obtaining $\lambda F_{\lambda} \propto \lambda^{-0.67 \pm 0.01}$. This slope is significantly shallower than the theoretical expectation, indicating a deviation from the standard accretion disk prediction.

To better isolate the variable component of the emission, we constructed an alternative SED using the difference between the maximum and minimum flux states ($\Delta F_{\lambda}$) of the light curves \citep[also see;][]{2016ApJ...821...56F} . This difference SED, shown in the middle panel of Figure \ref{fig:fvar}, provides a more accurate representation of the intrinsic variability of the accretion disk while effectively minimizing contamination from host galaxy starlight in different filters. Fitting this differential SED yields a steeper slope of $\lambda F_{\lambda} \propto \lambda^{-1.52 \pm 0.05}$, which is in much closer agreement with the $-4/3$ slope predicted by the thin disk model. For reference, we also overplot the theoretical SED slope of $-4/3$ on both panels by fixing the normalization constant accordingly. This facilitates a direct visual comparison between the observed trends and the canonical thin accretion disk expectation.

In the bottom panel of Figure~\ref{fig:fvar}, we show $F_{\mathrm{var}}$ as a function of wavelength. Across all five filters used in this study, we observe significant flux variability, with $F_{\mathrm{var}}$ values ranging from 0.0868 to 0.1542. Notably, 
$F_{\mathrm{var}}$ shows a gradual decrease with increasing wavelength in Fairall~9, consistent with the commonly observed 'bluer when brighter' trend in AGNs \citep{2010ApJ...711..461S}. However, this decreasing trend flattens at longer wavelengths, suggesting increased contamination from host galaxy starlight, which effectively dilutes the observed variability in those bands (also see Section~\ref{host-sub} for host-galaxy subtraction). Furthermore, the outer regions of the accretion disk, which emit primarily at optical and near-infrared wavelengths, act as a low-pass filter due to their longer dynamical and thermal timescales. This filtering smooths out rapid fluctuations, thereby reducing the amplitude of variability. As a result, the observed $F_{\mathrm{var}}$ is both diluted and damped at longer wavelengths.

\subsection{Host subtracted AGN luminosity}
\label{host-sub}
\begin{figure}
\resizebox{9.2cm}{6.0cm}{\includegraphics{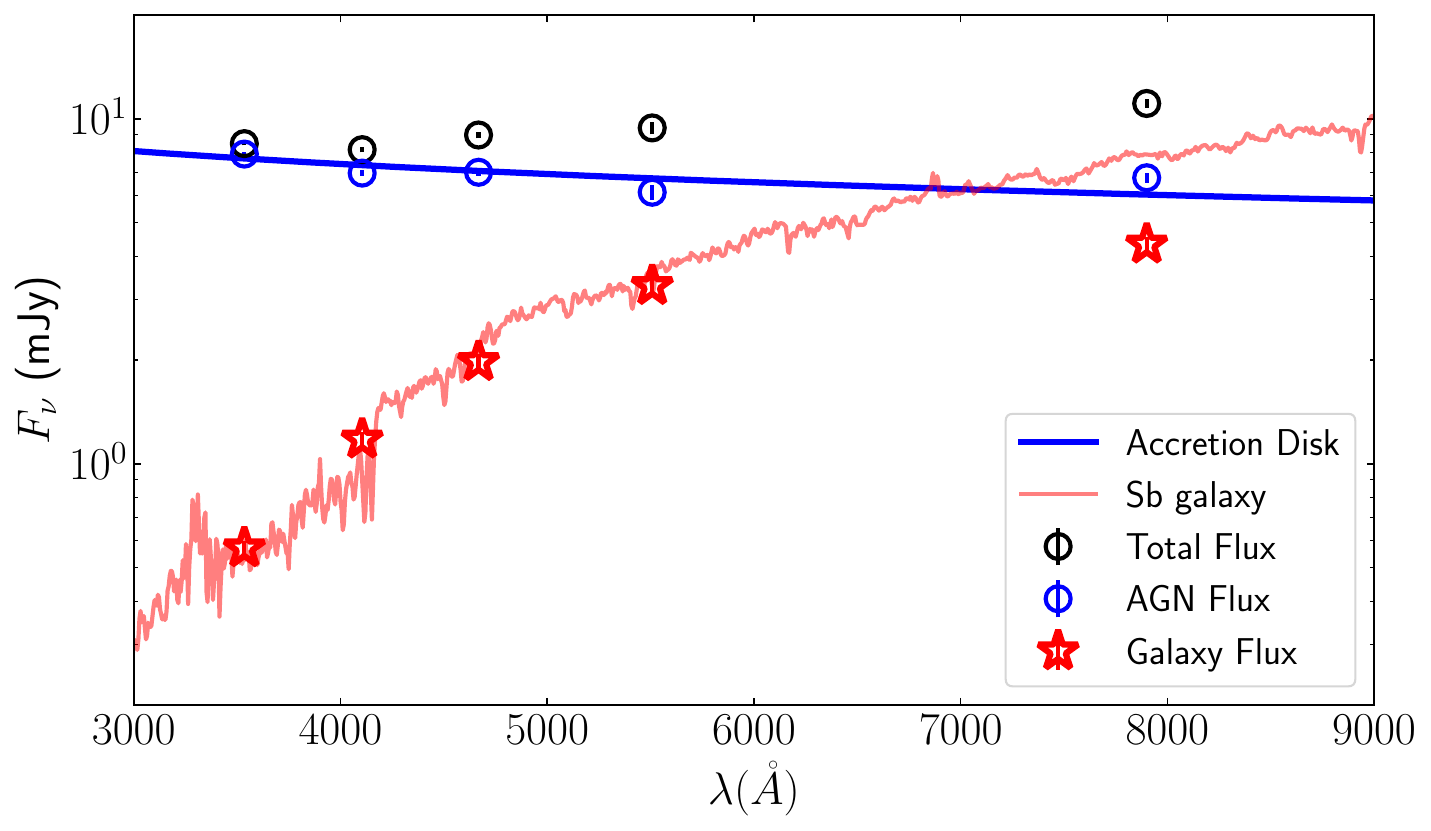}}

\caption{Spectral flux decomposition obtained with the FVG method. Black circles show the total flux density $F_\nu$. The host-subtracted AGN spectrum (blue circles) follows the thin accretion-disk prediction F$_{\nu}\propto\nu^{1/3}$ (\citealt{SS1973}). The host component inferred from the FVG (red stars) is well matched by an Sb galaxy template (red curve; \citealt{1996ApJ...467...38K}).}
\label{fig:median_sed}
\end{figure}

To estimate the contribution of the host galaxy and isolate the AGN emission, we applied the Flux Variation Gradient (FVG) method as described in \cite{pozo2012}. 
In essence, the method assumes that, for a fixed photometric aperture, flux measurements in different filters are the sum of a variable AGN component and a constant host-galaxy component, the latter including emission from the narrow-line region. 
Although the AGN flux varies over time, its optical spectral shape remains effectively constant. 
Consequently, fluxes measured in two bands define a linear relation in flux$-$flux space, where the slope corresponds to the AGN color and the host galaxy color is confined to a well-defined range (e.g. \citealt{2010ApJ...711..461S}). 
The intersection of the AGN slope with the host-galaxy color range then yields the host flux in each band for the period of monitoring.

Because light curves are affected by photometric noise, we estimated the intersection region using the density of possible solutions from a Bayesian principal component analysis \citep{2022A&A...657A.126G}. 
Figure~\ref{fig:fvg} in Section~\ref{ap:fvg} shows the FVG diagrams for all filter pairs, along with the adopted host-galaxy color ranges from \cite{2010ApJ...711..461S}. 
The resulting median values of AGN, and host-galaxy fluxes for each filter are listed in Table \ref{tab:var}.

Our results show that the fractional contribution of the host galaxy increases steadily with wavelength, in agreement with expectations for stellar-dominated emission at longer wavelengths. 
In the bluest $u$ band, the host accounts for only $\sim 7\%$ of the total flux, rising to $\sim 14\%$ in $v$ and $\sim 22\%$ in $b$. 
Beyond $5000$~\AA, the host fraction becomes more significant, reaching $\sim 35\%$ in $y$ and nearly $40\%$ in the $I$ band (see Table~\ref{tab:var}). This wavelength dependence reflects the increasing prominence of the old stellar population in the host galaxy, whose SED peaks in the optical–NIR, while the AGN continuum remains comparatively blue. Accurate host subtraction is therefore particularly critical in the redder filters, where stellar contamination can dominate the flux budget. 

Figure~\ref{fig:median_sed} presents the median SED after decomposing the total flux into AGN and host-galaxy components using the FVG method. When compared with an Sb galaxy template scaled to the inferred host fluxes, the host contribution shows close agreement across the optical bands. Although \citet{2009ApJ...697..160B} reported an SBa-type morphology for Fairall~9, our results favor an intermediate-type spiral stellar population as the dominant contributor to its host galaxy. 
The $I$-band point lies modestly below the Sb template prediction. Given the low photometric uncertainties and the consistent extraction aperture across all bands, this offset is unlikely to be due to measurement or calibration effects. 

A more plausible explanation is that the host stellar population differs slightly from that of a canonical Sb galaxy, with a reduced contribution from late-type giants, leading to a lower near-infrared flux. This is consistent with the morphological classification suggested by \citet{Jiang2021} who, from visual inspection of the \textit{Hubble} ACS image presented by \citet{2009ApJ...697..160B}, proposed a likely SB0/a type. Such an intermediate morphology could produce an SED broadly similar to Sb in the optical while exhibiting reduced flux at longer wavelengths. Nevertheless, the consistency between the difference SED shown in Figure~\ref{fig:fvar} and the host-subtracted SED presented in Figure~\ref{fig:median_sed} with the thin accretion disk prediction further supports the reliability of our flux decomposition based on the FVG method.

\subsection{Detrending of light curves}
\label{ss:detrend}

In addition to the rapid variations associated with short-timescale accretion disk reverberation, the observed UV-optical light curves also exhibit slower variability on longer timescales. These longer timescales are related to two separate phenomena. The first is the stochastic, multi-scale variability of the accretion rate in the innermost regions of the flow \citep[e.g.][]{vaughan2003}, most likely arising from local fluctuations that propagate inward \citep{lyubarski1997}. Such variability can give rise to long-timescale negative delays, which have now been tentatively identified in some sources \citep[e.g.][ specifically in Fairall~9]{Secunda2023}. The second phenomenon is the contribution of reprocessing from more distant regions such as the BLR, and torus, which leads to positive time delays.  To isolate the rapid variability and recover short lags, a detrending method is often applied by subtracting the slower variability component from the observed UV-optical light curves \citep{2010ApJ...721..715D, 2018ApJ...869..137L, 2018MNRAS.480.2881M, 2020A&A...642A..59R, 2020ApJ...896..146Z, 2024ApJ...962...67W, edelson2024, 2024A&A...683A.140Z}. For instance, \cite{edelson2024} detrended their light curves by fitting and subtracting a second-order polynomial, thereby removing long-term trends.

Following this approach, we adopted the methodology of \cite{edelson2024} for detrending our light curves. Specifically, we fit the entire light curves with polynomials of order 0, 1, and 2, where order = 0 corresponds to the original, unmodified light curves. We did not employ polynomial orders higher than 2, since higher-order detrending risks oversubtracting genuine short-timescale variability \citep{edelson2024} and introducing spurious artifacts into the light curves.

In our subsequent analysis, we therefore made use of both the original and detrended light curves. We emphasize, however, that there is no universally accepted choice for the polynomial order in detrending, and in all cases this process carries the risk of generating artificial variability in the light curves (for more details see the discussion in Section~\ref{ss:eff_detrend} and Figure~\ref{fig:detrend_lc}). For this reason, we treat the original light curves as our primary dataset, while using the detrended versions only for comparison.

\subsection{Lag analysis}
\label{lag_analysis}

During our monitoring period, Fairall~9 exhibited correlated significant flux variations across all the five filters used. Therefore, to measure the inter-band time delays, we employed the Interpolated Cross-Correlation Function (ICCF) method as our primary analysis tool, implemented via the PyCCF python package \citep[ICCF;][]{1986ApJ...305..175G, 1987ApJS...65....1G, 1998PASP..110..660P, 2018ascl.soft05032S}. This method involves linearly interpolating the two light curves and computing the cross-correlation coefficient (r). The uncertainties are estimated using a model-independent Monte Carlo simulation based on flux randomization (FR) and random subset selection (RSS) method following \citet{1998PASP..110..660P, 1999ApJ...526..579W}; and  \citet{peterson2004}. We repeated the simulation for 5000 iterations, with the lag centroid calculated in each iteration. Subsequently, these centroids were used to construct the cross-correlation centroid distribution (CCCD). For each iteration, the centroid lag was determined using only the points above $80\%$ of the peak value of the cross-correlation function ($r_{max}$). The final lag value was obtained from the median of the CCCD, while the associated uncertainties were defined by the 15.9th and 84.1st percentiles of the distribution corresponding to a 1$\sigma$ confidence interval assuming a Gaussian distribution. We used the $u$ band as the reference light curve, as it corresponds to the shortest wavelength in our monitoring program, and measured the inter-band delays between $u$ versus $v$ ($\tau_{uv}$), $u$ versus $b$ ($\tau_{ub}$), $u$ versus $y$ ($\tau_{uy}$), and $u$ versus $I$ ($\tau_{uI}$).

\begin{figure*}
    \centering
   \includegraphics[width=0.3\linewidth]{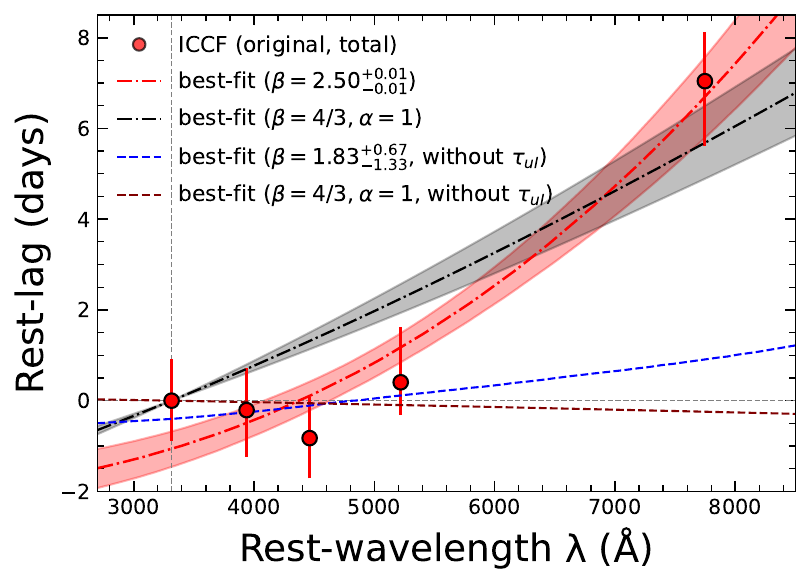}
   \includegraphics[width=0.3\linewidth]{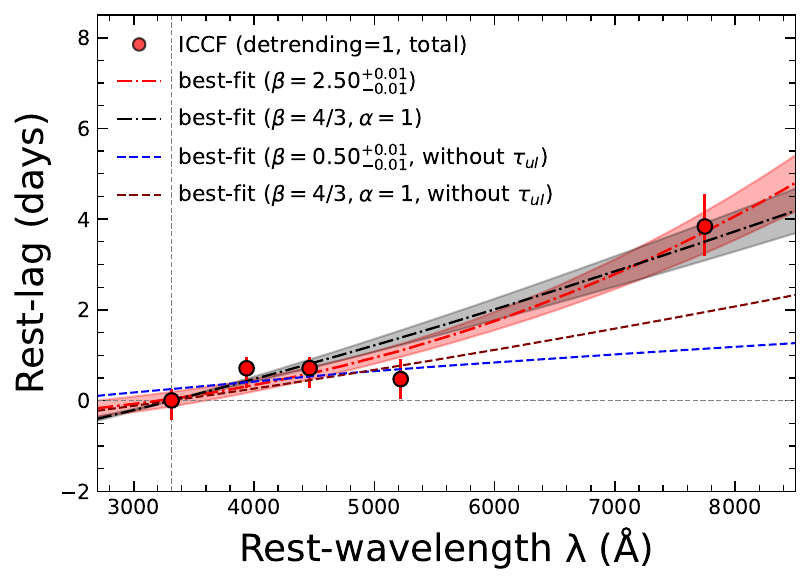}
   \includegraphics[width=0.3\linewidth]{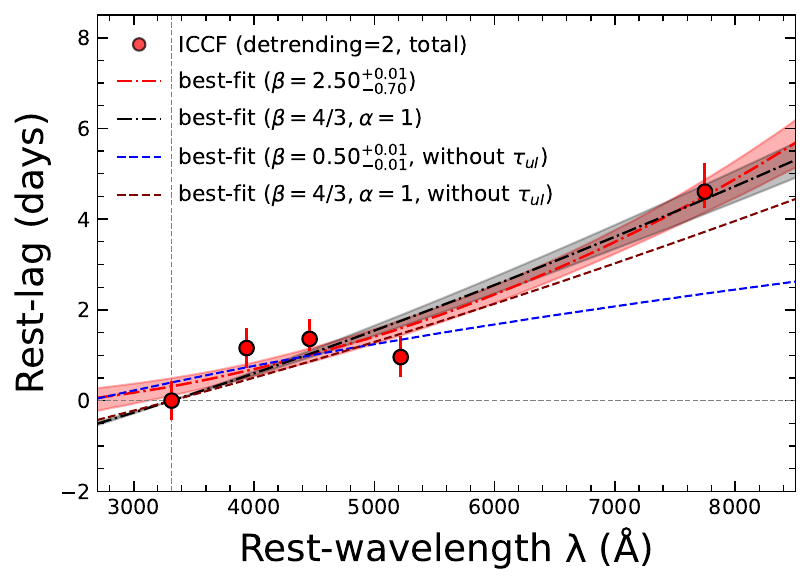}
    \caption{Examples of lag-spectra based on ICCF lag measurements using the complete light curves for three cases: original (left), detrended with order = 1 (middle), and detrended with order = 2 (right). The ICCF lag measurements (red circles) are shown in the rest-frame as a function of rest-frame wavelength. The red dot-dashed line represents the best-fit to the relation $\tau \, \propto \lambda^{\beta}$ using Equation~\ref{fit_eq}, where $\tau_{0}$, $\beta$, and $\alpha$ are free parameters. The black dot-dashed line shows the best-fit with fixed $\beta=4/3$ and $\alpha=1 $ with only $\tau_{0}$  as a free parameter. Shaded regions indicate the 1$\sigma$ uncertainties from the fitting. The blue and maroon dashed lines show the respective best-fits derived without including $\tau_{uI}$ in the lag-spectrum, for the cases where all parameters ($\tau_{0}$, $\beta$, and $\alpha$) are free, and where $\beta = 4/3$ and $\alpha=1$ are fixed. Vertical and horizontal dotted lines mark the rest-frame reference wavelength and zero rest-lag, respectively.}
    \label{fig:disk_map}
\end{figure*}

In addition to the ICCF method, we also employed {\tt PyROA} \citep{2021MNRAS.508.5449D} and {\tt JAVELIN} \citep{2011ApJ...735...80Z} to measure inter-band time delays as a consistency check. {\tt PyROA} models AGN light curve variability using the running optimal average (ROA) technique, where model parameters are sampled via Markov Chain Monte Carlo (MCMC) methods. This empirical approach does not rely on the commonly assumed damped random walk \citep[DRW;][]{kelly2009, 2010ApJ...721.1014M} model. The effective number of parameters in the ROA model is governed by the shape and width ($\Delta$) of its window function, as well as by the sampling cadence and precision of the data. This setup allows for a Bayesian optimization of the ROA width and other parameters, using Bayesian information criterion (BIC) minimization and MCMC sampling \citep{2021MNRAS.508.5449D}. Ultimately, {\tt PyROA} provides an estimate of the time delay between light curves at different wavelengths.

{\tt JAVELIN} models and interpolates AGN light curves using the DRW framework \citep{2011ApJ...735...80Z}. First, it fits the reference light curve with a DRW model characterized by two parameters, the variability amplitude ($\sigma_d$) and the time scale of variability ($\tau_d$). The reprocessed continuum light curve is then generated by convolving this model with a top-hat transfer function. Next, the time lag is estimated by maximizing the likelihood through MCMC sampling, with the final lag taken as the median of the resulting lag probability distribution.

The time delays derived from the three methods, based on various combinations of light curve data both with and without detrending, are summarized in Section~\ref{ap:lag} and presented in Table~\ref{tab:lag}.

Comparing different methods, we notice small inconsistencies between the nonparametric, model-independent method ICCF and model-dependent approaches that rely on Bayesian or stochastic modeling based on the DRW process, such as {\tt PyROA} and {\tt JAVELIN}. While ICCF tends to yield more conservative uncertainties and remains robust even in the presence of flux error mis-estimates, both {\tt PyROA} and {\tt JAVELIN} are more sensitive to the underlying modeling assumptions. These assumptions may not fully capture the true nature of AGN variability, particularly when attempting to model short-timescale fluctuations in light curves that span long temporal baselines.

For instance, {\tt PyROA} fails to accurately model the short-timescale rapid variations, specifically in the $b$, $y$ and $I$ light curves (as example see Figure \ref{fig:lc} for the period MJD $\sim$60675 to 60715), leading to significant deviations in the inferred lags, $\tau_{uy}$ and $\tau_{uI}$, when compared with those obtained from ICCF. Similar discrepancies are also evident when using {\tt JAVELIN}. This issue is further supported by earlier studies, such as, \citet{2025ApJ...985...30M} reported that {\tt JAVELIN} often fails to recover accurate lag values in the presence of large seasonal gaps in the data, and \citet{2024ApJS..275...13W} suggested that ICCF offers the most robust lag measurements compared to {\tt PyROA} and {\tt JAVELIN}. Furthermore, we compare the lags obtained from these three methods in Figure~\ref{lag_comp} and discuss them in detail in Section~\ref{ss:lag_compare}.

Given these limitations, and despite the larger associated uncertainties, we adopt the lags obtained from ICCF as our final measurements in the subsequent analysis, prioritizing reliability and robustness over potentially underestimated errors from model-dependent methods. Note that the inclusion or exclusion of the single data point at MJD~60573.238, which is separated by a large observational gap in the $I$ light curve, does not affect the derived lags. Therefore, we included all available data points in the $I$ band for our analysis.

\section{Results}
\label{res}

\subsection{Disk-mapping: Traditional method with $\tau_{\lambda} \, \propto \, \lambda^{\beta}$}

According to the lamp-post model, an X-ray emitting corona illuminates the accretion disk, whose reprocessed UV-optical emission shows delayed responses to variations in the shorter-wavelength continuum. This delay arises from the radial temperature gradient in the disk, $T(r) \, \propto \,  r^{-3/4}$, as predicted by the standard accretion disk model. Consequently, the size of the standard disk at wavelength $\lambda$ can be expressed as \citep{2017ApJ...840...41E}
\begin{equation} \label{ssd_eq}
R_{\lambda,\text{SSD}} = \left(\chi\frac{\kappa \lambda}{hc}\right)^{4/3} \left[\left(\frac{GM_{\text{BH}}}{8 \pi \sigma}\right) \left(\frac{L_{\text{Edd}}}{\eta c^2}\right) (3+\kappa)\lambda_{\text{Edd}}\right]^{1/3},
\end{equation}
where $G$ is the gravitational constant, $M_{\text{BH}}$ the black hole mass, $\sigma$ the Stefan–Boltzmann constant,  $L_{\text{Edd}}$ the Eddington luminosity, and  $\lambda_{\text{Edd}}$ the Eddington ratio. The parameter $\chi$ is a scaling factor that accounts for systematic discrepancies in converting the annulus temperature $T$ to a wavelength $\lambda$ at a characteristic radius. The radiative efficiency is denoted by $\eta$, and $\kappa$ represents the ratio of external to internal disk heating. In this work, we assume $\eta$ = 0.1 and  $\kappa$ = 1, corresponding to equal contributions from X-ray and viscous heating. We adopt $\chi$ = 2.49, consistent with a flux-weighted radius \citep{edelson2024}, while Wien’s law predicts a larger value of  $\chi = 4.97$ \citep{netzer2022}. Note that there is no standard value of $\chi$, which introduces additional uncertainty in Equation \ref{ssd_eq}.

As a result, the inter-band continuum time delays follow a power-law dependence on wavelength as $\tau_{\lambda} \, \propto \, \lambda^{4/3}$. Measuring these continuum reverberation lags across different UV–optical bands therefore provides a way to infer the temperature profile of the disk. This technique, referred to as disk-mapping, is implemented by fitting the observed inter-band delays as a function of wavelength using the relation

\begin{equation} \label{fit_eq}
\tau_{\lambda} =  \tau_{0} \left[\left(\frac{\lambda}{\lambda_{0}}\right)^{\beta}-\alpha\right],
\end{equation}
where $\lambda_{0}$ represents the reference-band wavelength,  $\tau_{0}$ is the normalization amplitude of the power-law lag-spectrum at $\lambda_{0}$, and $\beta$ is the power-law index, which is $\sim$ 4/3 as prescribed by the SSD model. The parameter $\alpha$ introduces flexibility by allowing the model lag at $\lambda_{0}$ to differ from zero; specifically, $\alpha = 1$ forces the lag at $\lambda_{0}$ to be zero. Hence, the normalization constant $\tau_{0}$ inferred from disk-mapping provides a measure of the accretion disk size at $\lambda_{0}$, given by $R_{\lambda_{0}} = c \times \tau_{0}$.

Although many studies have reported that inter-band delays broadly follow the expected power-law behaviour, $\tau_{\lambda} \, \propto \lambda^{4/3}$ \citep{2022ApJ...940...20G, 2023ApJ...947...62K, edelson2024, 2025ApJ...985...30M, 2025A&A...700L...8P}, significant departures from this relation have also been observed in numerous cases \citep{2016ApJ...821...56F, 2017ApJ...835...65S, 2023A&A...672A.132F, 2023MNRAS.519.3366M, 2025MNRAS.542.2572G}. Furthermore, disk sizes inferred from disk-mapping are typically found to be about 4$-$5 times ($0.69\pm0.04$ dex) larger than those predicted by the standard SSD model, a discrepancy commonly referred to as the 'disk-size anomaly' \citep{cackett_2021iSci, 2025ApJ...985...30M}. Similar disk-size anomalies have also been detected in microlensing studies of gravitationally lensed quasars  \citep{rauch1991, 2010ApJ...712.1129M, 2011ApJ...729...34B, 2016AN....337..356C, 2019MNRAS.483.2275L, 2024SSRv..220...14V}.

\subsubsection{For HALO dataset}

Motivated by these findings, we first perform disk-mapping on Fairall~9 using data from our HALO program, assuming a power-law dependence of $\tau_{\lambda}$ on wavelength $\lambda$ using Equation~\ref{fit_eq}, with $\tau_{0}$, $\beta$, and $\alpha$ as free parameters, in order to examine any potential departure from the SSD-predicted slope. We then repeat the analysis by fixing the slope to $\beta_{\text{fix}} = 4/3$, as expected from the SSD model, and setting $\alpha$ to 1, thereby forcing the lag to be zero at $\lambda_{0}$.

In Figure~\ref{fig:disk_map}, we present the lag-spectra, i.e., the rest-frame inter-band time delays as a function of rest-frame wavelength, using the complete light curves for three different cases: the original light curves, and detrended light curves of order 1 and order 2, shown in the left, middle, and right panels, respectively. When we treat $\tau_{0}$, $\beta$, and $\alpha$ as free parameters, we find that the resulting power-law slope varies significantly, ranging between 0.91 and 2.50 for different data combinations, and thus deviates considerably from the SSD-predicted value of 4/3.

However, the disk sizes inferred from the normalization parameter $\tau_{0}$ at the reference wavelength $\lambda_{0}$ cannot be directly compared with the SSD-predicted value, since $\beta$ deviates significantly from $4/3$, and the $\tau = 0$ lags in the best-fit models are shifted relative to the reference wavelength $\lambda_{0} = 3468.0\ \text{\AA} / (1 + z)$.

When we repeat the analysis by fixing $\beta$ and $\alpha$ to 4/3 and 1, respectively, leaving only $\tau_{0}$ as a free parameter, we find systematically larger disk sizes, about 1.5$-$3.6 times greater than the SSD prediction. These results consistently indicate a significant departure from the SSD expectations.

Since the object variability shows an apparent change in the middle of our monitoring (see Figure~\ref{fig:lc}), we also analyzed separately the two segments: S1 and S2, which are divided by an observational gap. The best-fit values obtained from our disk-mapping analysis are summarized in Table~\ref{tab:disk_size}, where the corresponding $\chi^2$ per degree of freedom ($\chi^2/\text{dof}$) values are also listed. Among the various disk size measurements derived from different data combinations, the original dataset for S2 and the detrended dataset with order = 1 for S1 provide comparatively better estimates, with $\chi^2/\text{dof} = 0.98\ (0.78)$ and $1.00\ (1.25)$ when $\beta$ and $\alpha$ are treated as free (fixed) parameters during fitting, respectively. The corresponding disk sizes at $\lambda_{0} = 3468.0\ \text{\AA}/(1+z)$ are  $R_{3315} = 1.62_{-1.17}^{+3.80}\ (1.52_{-0.17}^{+0.17})$  light-days for S2-original and  $0.68_{-0.21}^{+0.26}\ (1.83_{-0.57}^{+0.56})$  light-days for S1-detrending=1.

For comparison, \citet{hernandez2020} and \citet{edelson2024} reported disk sizes of $R_{1928} = 1.20 \pm 0.10$ light-days, and $R_{2050} = 1.81 \pm 0.20$ light-days at reference wavelengths of 1928 {\AA} and 2050 {\AA}, respectively. When extrapolated to 3315 {\AA} with $\beta_{\text{fix}} = 4/3$, these correspond to $R_{3315} = 2.47 \pm 0.21$ light-days, and $3.44 \pm 0.38$ light-days, respectively. In contrast, our estimated disk sizes are smaller by a factor of $\sim 2-4$ compared to those reported by \citet{hernandez2020} and \citet{edelson2024}. This discrepancy can arise due to (i)  firstly, the inter-band delays do not strictly follow a simple power-law; instead, the lag-spectrum exhibits wavelength-dependent wiggles. These features are primarily caused by Balmer continuum contamination in the $u$ (or Swift $U$, see the following section) band and diffuse continuum emission due to BLR reprocessing across all continuum-tracing filters. Consequently, disk size estimates based on power-law fits may vary between monitoring campaigns, depending on the wavelength coverage. (ii) Secondly, differences in data quality between monitoring periods can further affect the measured lags and, in turn, the resulting lag-spectrum fits. Our observations, obtained with intermediate-band Str\"omgren filters, benefit from denser sampling ($\Delta t \sim 0.3$ days) and the absence of emission-line contamination, thereby offering a clearer recovery of subtle accretion-disk reverberation signals. In contrast, their campaigns relied on broad-band filters with relatively sparse sampling ($\Delta t \sim 1.3$ days), making them more susceptible to emission-line dilution and less sensitive to small-scale disk variations. And (iii) thirdly, the small variations in disk sizes estimated from different seasons (S1 and S2) of our light-curve observations further suggest that the source may undergo intrinsic changes in the accretion disk-reverberation timescale across epochs.

\begin{table}[h!]
\centering

 \caption{Parameters estimated from the disk-mapping}
 \label{tab:disk_size}

\resizebox{9cm}{!}{
\fontsize{20pt}{20pt}\selectfont
\begin{tabular}{ccccc} \hline \hline

Data & $\tau_{0}$ & $\beta$ & $\alpha$  & $\chi^2$/dof 
\\ 
 & (days) &  &  & \\
(1) & (2) & (3) & (4) & (5)
\\ \hline \\
Original-total & $1.06_{-0.14}^{+0.14}$ & $2.50_{-0.01}^{+0.01}$ & $2.00_{-0.32}^{+0.31}$ & 3.21 \\

Original-total & $2.72_{-0.37}^{+0.37}$ & $4/3$ & 1 & 5.57 \\ 

 Original-total (without $\tau_{uI}$) & $0.27_{-0.19}^{+0.47}$ & $1.83_{-1.33}^{+0.67}$ & $2.42_{-1.53}^{+1.58}$ & 1.76 \\

 Original-total (without $\tau_{uI}$) & $-0.13_{-0.68}^{+0.69}$ & $4/3$ & 1 & 0.68 \\ \\

Detrending=1-total & $0.51_{-0.07}^{+0.09}$ & $2.50_{-0.01}^{+0.01}$ & $0.93_{-0.34}^{+0.29}$ & 3.50 \\

Detrending=1-total & $1.66_{-0.19}^{+0.20}$ & $4/3$ & 1 & 2.62 \\

Detrending=1-total (without $\tau_{uI}$) & $1.58_{-1.12}^{+1.57}$ & $0.50_{-0.01}^{+0.01}$ & $0.83_{-0.83}^{+0.15}$ & 3.63 \\

Detrending=1-total (without $\tau_{uI}$) & $0.93_{-0.29}^{+0.31}$ & $4/3$ & 1 & 1.85 \\ \\

Detrending=2-total  & $0.61_{-0.07}^{+0.54}$ & $2.50_{-0.70}^{+0.01}$ & $0.52_{-0.43}^{+0.39}$ & 5.11 \\

Detrending=2-total & $2.11_{-0.15}^{+0.16}$ & $4/3$ & 1 & 3.01 \\

Detrending=2-total (without $\tau_{uI}$) & $3.62_{-1.62}^{+1.67}$ & $0.50_{-0.01}^{+0.01}$ & $0.89_{-0.21}^{+0.08}$ & 7.15 \\

Detrending=2-total (without $\tau_{uI}$) & $1.76_{-0.30}^{+0.30}$ & $4/3$ & 1 & 3.79 \\ \\

Original-S1 & $0.55_{-0.20}^{+4.69}$ & $2.43_{-1.93}^{+0.07}$ & $1.15_{-0.70}^{+1.18}$ & 1.30 \\

Original-S1 & $1.19_{-0.30}^{+0.30}$ & $4/3$ & 1 & 0.15 \\ 

Original-S1 (without $\tau_{uI}$) & $0.65_{-0.49}^{+2.45}$ & $0.50_{-0.01}^{+2.00}$ & $1.14_{-1.14}^{+2.53}$ & 0.28 \\

Original-S1 (without $\tau_{uI}$) & $0.57_{-1.07}^{+1.06}$ & $4/3$ & 1 &  0.01 \\ \\

Detrending=1-S1 & $0.68_{-0.21}^{+0.26}$ & $2.50_{-0.06}^{+0.01}$ & $1.44_{-0.48}^{+0.46}$ & 1.00 \\

Detrending=1-S1 & $1.83_{-0.57}^{+0.56}$ & $4/3$ & 1 & 1.25 \\

Detrending=1-S1 (without $\tau_{uI}$) & $0.47_{-0.33}^{+2.39}$ & $0.50_{-0.01}^{+2.00}$ & $1.01_{-1.01}^{+0.72}$ & 0.72 \\

Detrending=1-S1 (without $\tau_{uI}$) & $0.55_{-0.57}^{+0.62}$ & $4/3$ & 1 & 0.08 \\ \\

Detrending=2-S1 & $0.48_{-0.15}^{+0.16}$ & $2.50_{-0.01}^{+0.01}$ & $1.82_{-0.36}^{+0.32}$ & 2.93 \\

Detrending=2-S1 & $1.28_{-0.42}^{+0.42}$ & $4/3$ & 1 & 3.69 \\

Detrending=2-S1 (without $\tau_{uI}$) & $0.11_{-0.08}^{+0.45}$ & $0.50_{-0.01}^{+2.00}$ & $1.28_{-1.27}^{+2.72}$ & 0.08 \\

Detrending=2-S1 (without $\tau_{uI}$) & $0.01_{-0.35}^{+0.35}$ & $4/3$ & 1 & 0.01 \\ \\

Original-S2 & $1.62_{-1.17}^{+3.80}$ & $1.20_{-0.70}^{+1.22}$ & $0.89_{-0.86}^{+0.11}$ & 0.98 \\

Original-S2 & $1.52_{-0.17}^{+0.17}$ & $4/3$ & 1 & 0.78 \\

Original-S2 (without $\tau_{uI}$) & $3.82_{-2.92}^{+2.05}$ & $0.50_{-0.01}^{+0.16}$ & $0.93_{-0.25}^{+0.07}$ & 1.73 \\

Original-S2 (without $\tau_{uI}$) & $1.68_{-0.34}^{+0.36}$ & $4/3$ & 1 & 0.95 \\ \\

Detrending=1-S2 & $1.84_{-1.48}^{+2.27}$ & $0.91_{-0.41}^{+1.43}$ & $0.88_{-0.86}^{+0.11}$ & 1.34 \\

Detrending=1-S2 & $1.16_{-0.16}^{+0.17}$ & $4/3$ & 1 & 1.28 \\

Detrending=1-S2 (without $\tau_{uI}$) & $2.97_{-1.93}^{+1.60}$ & $0.50_{-0.01}^{+0.01}$ & $0.92_{-0.24}^{+0.07}$ & 2.63 \\

Detrending=1-S2 (without $\tau_{uI}$) & $1.36_{-0.30}^{+0.31}$ & $4/3$ & 1 & 1.39 \\ \\

Detrending=2-S2 & $0.68_{-0.16}^{+2.96}$ & $2.17_{-1.32}^{+0.33}$ & $0.60_{-0.60}^{+0.39}$ & 3.20 \\

Detrending=2-S2 & $1.96_{-0.19}^{+0.17}$ & $4/3$ & 1 & 1.84 \\

Detrending=2-S2 (without $\tau_{uI}$) & $4.04_{-1.92}^{+1.94}$ & $0.50_{-0.01}^{+0.01}$ & $0.91_{-0.21}^{+0.08}$ & 4.06 \\

Detrending=2-S2 (without $\tau_{uI}$) & $1.86_{-0.33}^{+0.33}$ & $4/3$ & 1 & 2.42 \\ \\ \hline \\

HALO+Literature & $1.66_{-0.53}^{+1.85}$ & $2.10_{-0.76}^{+0.40}$ & $1.87_{-0.45}^{+0.37}$ & 11.51 \\

HALO+Literature & $2.77_{-0.33}^{+0.32}$ & $4/3$ & 1 & 39.30 \\

HALO+Literature (without $\tau_{uI}$) & $7.37_{-1.64}^{+1.68}$ & $0.50_{-0.01}^{+0.01}$ & $1.16_{-0.03}^{+0.04}$ & 3.93 \\

HALO+Literature (without $\tau_{uI}$) & $1.65_{-0.45}^{+0.47}$ & $4/3$ & 1 & 15.79 \\ \\

\hline

\end{tabular}
}

\tablefoot{Columns are (1) light curve data type used for lag analysis, (2) normalization constant at reference wavelength $\lambda_{\text{0}} = \mathrm{3468 \,\AA}/(1+z)$ from fitting, (3) power-law index, either free or fixed at $\beta_{\text{fix}} = 4/3$, (4) parameter, either free or fixed at 1, and (5) $\chi^2$ per degree of freedom. 
    }

\end{table}

\begin{figure}
    \centering
   \includegraphics[width=0.85\linewidth]{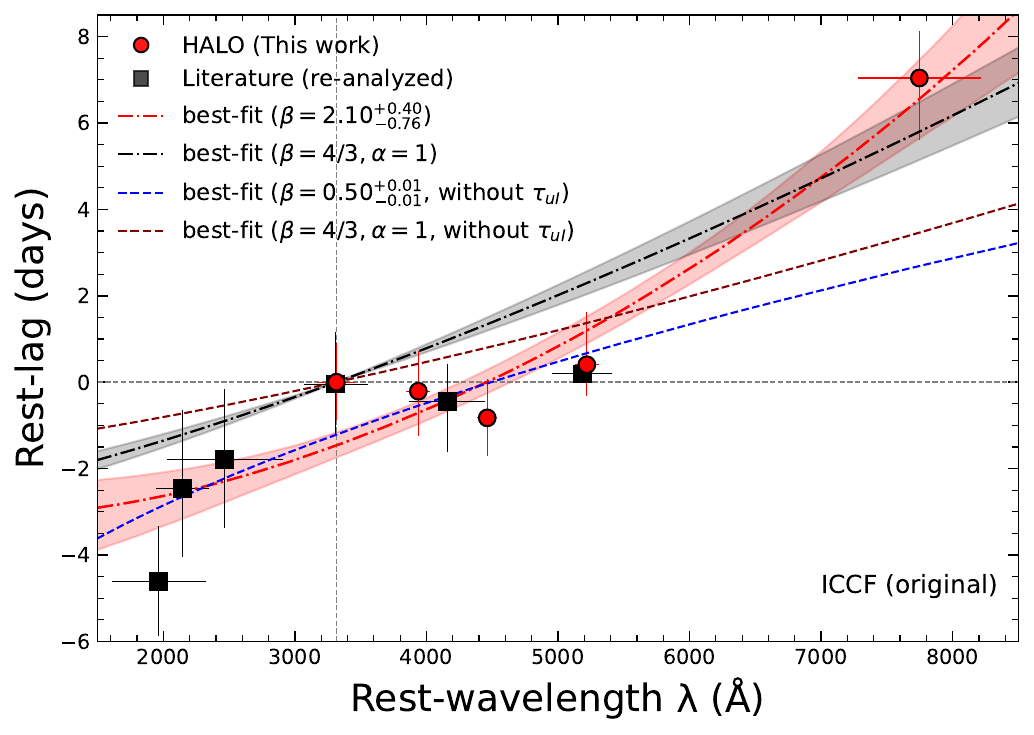}
    \caption{Combined lag–spectrum. Red circular points show our HALO measurements, while black square points indicate reanalyzed literature measurements based on the light curve data from \citet{edelson2024}. The horizontal error bars represent the rms width of the transmission curve as the filter’s wavelength uncertainty. The red and black dot-dashed lines, together with their shaded regions, represent the best-fit results from power-law lag–spectrum fitting, with $\beta$ and $\alpha$ treated as free parameters and fixed at $4/3$ and $1$, respectively. The corresponding fits obtained excluding $\tau_{uI}$ from the lag-spectrum are shown by the dashed blue and maroon lines. The lag–spectrum shows a clear excess in the $u$ and $U$ bands, suggesting Balmer continuum contamination from the BLR in these bands.}
    \label{fig:disk_map_tot}
\end{figure}

\subsubsection{For combined dataset: HALO + Literature}

To further investigate the discrepancies between our lag-spectra fitting of Fairall~9 and the results reported in the literature, we compiled additional light-curve data from literature. In particular, we made use of the Swift light curves assembled by \citet{edelson2024} (see Section~\ref{lit:lag} for details). The resulting delays are summarized in Table \ref{tab:lag_lit} and are hereafter referred to as the 'Literature' measurements.

We then plot the inter-band delays obtained from Swift observations together with our HALO measurements in the lag-spectrum shown in Figure~\ref{fig:disk_map_tot}. We exclude both HX- and SX-band lags from the lag-spectrum, as hard and soft X-rays are generally assumed to originate in the X-ray corona rather than from reprocessed emission in the accretion disk, leading to relatively weak correlations between the X-ray and UV-optical light curves. Remarkably, the lags within the overlapping wavelength range of filters show good consistency between measurements from HALO and those from Literature, despite derived from two independent monitoring programs carried out over distinct time periods. This consistency further suggests that the discrepancy in disk sizes inferred from these separate disk-mapping campaigns is not due to temporal variations or differences in data quality, but rather due to deviations from a simple power-law dependence of lag on wavelength. 

In addition, the combined lag-spectrum of Fairall~9 reveals a pronounced excess in the $u$ and $U$ bands, a feature that points to a significant contribution from the Balmer continuum in these bands. Such an excess has previously been observed in NGC~4593 \citep{cackett2018}, further supporting this interpretation.

These results highlight that the inter-band delays significantly deviate from the simple power-law trend $\tau_{\lambda} \, \propto \, \lambda^{\beta}$. In particular, when using the complete original light curve data, we find negative delays, $\tau_{uv}$, $\tau_{ub}$, and $\tau_{UB}$ which suggest a significant Balmer continuum contribution from the BLR to the $u$ and $U$ bands. However, when the light curves are detrended, the Balmer continuum contamination appears to diminish progressively with increasing detrending order. This reduction is likely due to the removal of long-timescale trends, which are expected to originate from the outer regions, such as BLR and/or torus. It should be noted, though, that there is no standard method for such detrending \citep{edelson2024} that avoids introducing artifacts into the observed light curves. Nevertheless, the results clearly indicate that the standard thin SSD model alone cannot account for the observed lag-spectra, since a simple power-law fit fails to reproduce individual features present in the AGN lag-spectra.

 However, $\tau_{uI}$ deviates significantly from the overall trend, exhibiting an excess lag in the lag-spectrum. This anomaly at the longest wavelength likely arises from contamination by the dust torus and the Paschen jump emission from the BLR (see Section~\ref{ss:rpc} for more details). Such excess lag can steepen the power-law slope, thereby affecting the normalization constant $\tau_{0}$. To assess the influence of $\tau_{uI}$ on the power-law fitting, we refitted the lag-spectrum after excluding $\tau_{uI}$; the best-fit results are listed in Table~\ref{tab:disk_size}. For the HALO dataset, the revised fit gives $\beta = 1.83$, which is lower than the $\beta = 2.50$ obtained from the lag-spectrum including $\tau_{uI}$ constructed using the original-total light curve data, yet remains steeper than the SSD prediction and is accompanied by a significantly smaller normalization constant. For other data combinations, we find even flatter lag–wavelength dependencies ($\beta \sim 0.5$), which makes it difficult to directly compare the normalization constant with cases that include $\tau_{uI}$.  However, when $\beta$ and $\alpha$ are fixed to 4/3 and 1, respectively, we obtain a negative disk size from the original-total light curves at a reference wavelength of 3315 {\AA} giving $R_{3315} = -0.13$ light-days. This result primarily arises from Balmer continuum contamination in the $u$ band, which causes both $\tau_{uv}$ and $\tau_{ub}$ to become negative. In all other cases, the observed $R_{3315}$ values are larger than the SSD-predicted value of 0.76 light-days, except in fits dominated by measurement uncertainties, where the reduced $\chi^2$/dof values ($<0.1$) are much smaller than unity. The similar behaviour is observed for HALO+Literature dataset. Therefore, the lag-spectra of Fairall~9, both with and without $\tau_{uI}$, show significant deviations from the SSD model predictions.

These findings motivate the development of a more comprehensive model that incorporates both accretion-disk emission and BLR reprocessing in order to explain the observed inter-band continuum time delays \citep{2025A&A...702A..92J}. A detailed treatment of such a model on Fairall~9 lag-spectrum including contamination from dust torus at longer wavelengths will be presented in a forthcoming paper (A.K. Mandal et al., in prep).

\section{Discussion}
\label{ss:dis}

The observed disk size anomalies, together with the departure from the simple power-law relation $\tau_{\lambda} \, \propto \, \lambda^{\beta}$ predicted by the standard thin-disk model with $\beta = 4/3$, remain central issues in the ongoing debate on AGN continuum lags. Recently, \citet{2025arXiv250925315L} classified AGNs into obscured and unobscured categories based on X-ray spectral modeling. They argued that the observed lag excess arises exclusively in obscured AGNs, whereas the continuum lags in unobscured AGNs are consistent with SSD predictions and remain unaffected by BLR contamination. However, our findings indicate that this claim does not hold universally. In particular, Fairall~9, despite being an unobscured AGN, clearly exhibits excess lags in the $u$ and $U$ bands within its lag-spectrum, attributable to BLR contamination, as demonstrated in this study. To account for these discrepancies further, several physical interpretations have been proposed. Early work \citep[e.g.,][]{Korista_Goad_2001, lawther2018} and subsequent observations \citep[e.g.,][]{2019NatAs...3..251C} established the BLR diffuse continuum as a key driver of inter-band lags. \citet{netzer2022} later proposed that inter-band time delays are dominated by diffuse emission from radiation-pressure-supported clouds in the BLR, while the contribution from an irradiated accretion disk is negligible. Notably, recent microlensing studies \citep[e.g.,][]{2023A&A...677A..94F, 2025A&A...695A..10H} also suggest that a diffuse BLR continuum contributes to the observed continuum signal, consistent with findings from continuum-RM. In contrast, \citet{2021ApJ...907...20K, kammoun2021} presented a fundamentally different view, suggesting that thermal reverberation within the accretion disk itself can explain the observed UV/optical time lags in AGN lag-spectra. More recently, \citet{2025A&A...702A..92J} demonstrated that the observed delays are likely produced by a combination of irradiated disk emission and additional contributions from the outer BLR through scattering and reprocessing, an interpretation further supported by the findings of \citet{pozo2023, 2025ApJ...985...30M}.

Motivated by these contrasting perspectives, in this section we tested both the radiation-pressure-confined (RPC) cloud model proposed by \citet{netzer2022} and the relativistic accretion disk model developed by \citet{kammoun2021} against the lag-spectrum of Fairall~9, in order to evaluate whether either framework could adequately reproduce the observed inter-band time delays.

\subsection{RPC model including BLR diffuse continuum and emission lines}
\label{ss:rpc}

\citet{netzer2022}, in his RPC model, proposed that the observed inter-band time delays are primarily driven by the diffuse continuum and emission line contributions from the BLR, while the intrinsic disk lag is so small that it remains unmeasurable with the resolution of current data. According to this model, the total observed continuum time delay ($\tau_{\mathrm{\lambda, tot}}$) at wavelength $\lambda$ can be expressed as,

\begin{equation}
\label{eq:rpc}
    \tau_{\mathrm{\lambda, tot}} = \tau_{\mathrm{\lambda,irr}}\bigg(\frac{L_{\mathrm{inc}}}{L_{\mathrm{inc}}+L_{\mathrm{diff}}}\bigg) +  \tau_{\mathrm{\lambda,diff}}\bigg(\frac{L_{\mathrm{diff}}}{L_{\mathrm{inc}}+L_{\mathrm{diff}}}\bigg),
\end{equation}
where $\tau_{\mathrm{\lambda,irr}}$ denotes the continuum lag from an irradiated disk, $\tau_{\mathrm{\lambda,diff}}$ represents the diffuse emission lag, $L_{\mathrm{inc}}$ is the combined luminosity of the accretion disk and central X-ray source, and $L_{\mathrm{diff}}$ includes the combined luminosity from broad emission lines, bound-free and free-free continua, as well as scattering by ionized and neutral gas.

\begin{figure}
    \centering
    \includegraphics[width=0.85\linewidth]{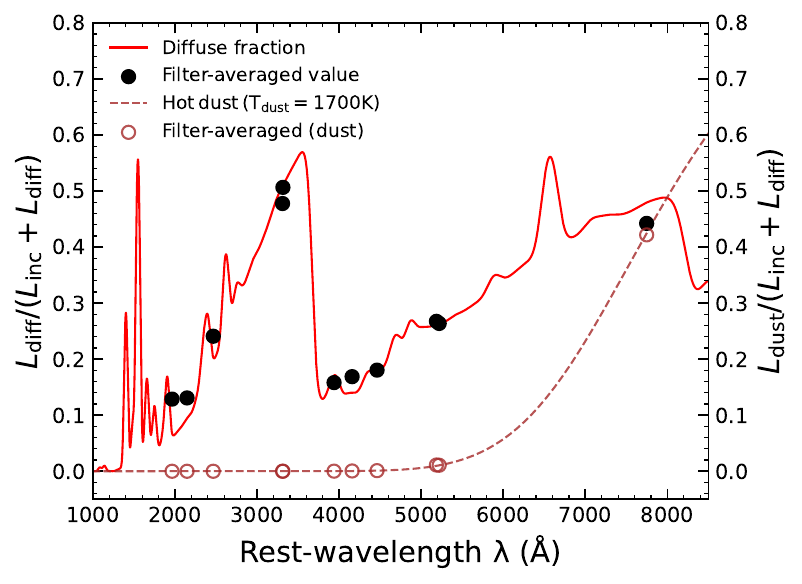}
   \includegraphics[width=0.85\linewidth]{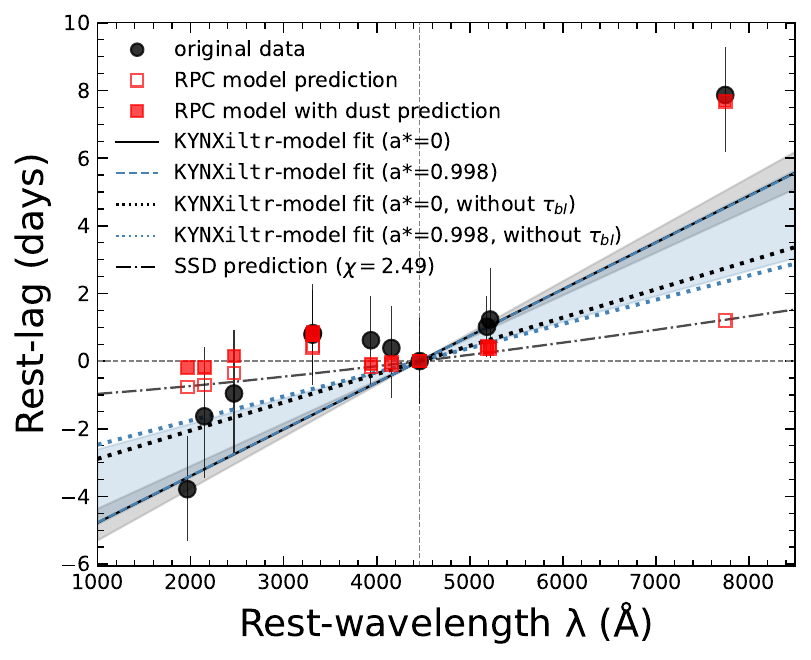}
    \caption{Lag-spectrum modeling. Top: Ratio $\frac{L_{\mathrm{diff}}}{L_{\mathrm{inc}}+L_{\mathrm{diff}}}$, calculated from the emissivity profile of BLR clouds centered at $\log r \, (\mathrm{cm}) = 16.65$ from {\tt Cloudy}, shown as a function of wavelength. Additionally, the dust fraction, $\frac{L_{\mathrm{dust}}}{L_{\mathrm{inc}}+L_{\mathrm{diff}}}$ , is overplotted as a dashed brown line. Black and brown points indicate the corresponding values averaged over the response curves of the filters used to measure continuum delays, respectively.
Bottom: Lag-spectrum of Fairall~9. Black points indicate the observed data; red open squares show the RPC-model recovered time delays; and solid red squares denote the RPC-model predictions when the contribution from hot torus dust is included in the continuum lags. Solid black and dashed blue lines represent the best-fit results from the {\tt KYNXiltr} model for spin $a\ast = 0$ and $a\ast = 0.998$, respectively. The corresponding fits obtained after excluding $\tau_{uI}$ from the lag-spectrum are shown by dotted lines. The fit ranges for the full lag-spectrum fitting are indicated by the gray and blue shaded regions.}
    \label{fig:model}
\end{figure}

Therefore, to apply the RPC model to the Fairall~9 data, we first reconstructed the lag-spectrum by calculating time delays relative to the $b$ band. This was achieved by subtracting $\tau_{ub}$ from all measured lags and propagating the corresponding errors. The choice of the $b$ band as reference was due to the fact that a power-law fit to the lag-spectrum with free parameters $\beta$ and $\alpha$ yielded a model time delay that approaches zero near this band (see Figure~\ref{fig:disk_map_tot}). Then, to determine the BLR spectral shape, we performed photoionization calculations with the {\tt Cloudy} code \citep{2023RMxAA..59..327C}. We adopted an incident bolometric luminosity of $L_{\text{bol}} = 10^{44.9} \, \mathrm{erg \, s^{-1}}$, estimated from $L_{5100}$ using a bolometric correction factor of 9.26 \citep{2006ApJS..166..470R}. The representative BLR cloud was assumed to be located at a distance of $\log r \, (\mathrm{cm}) = 16.65$ from the central source \citep[i.e., H$\beta$ BLR size;][]{peterson2004}, with a constant hydrogen gas density of $\log n_{\mathrm{H}} \, (\mathrm{cm^{-3}}) = 11$ and a column density of $\log N_{\mathrm{H}} \, (\mathrm{cm^{-2}}) = 23.5$ \citep{Korista_Goad_2001, 2022AN....34310091P}, while the metallicity was taken to be five times the solar value \citep{panda2019, naddaf2021, 2024A&A...689A.321F, 2025A&A...702A..92J}.

The resulting spectral shape of the ratio $\frac{L_{\mathrm{diff}}}{L_{\mathrm{inc}}+L_{\mathrm{diff}}}$ is shown in top panel of Figure~\ref{fig:model}. Then, we computed its filter-averaged values by convolving the model spectrum with the filter-response curves. These averages are shown as black points at representative rest-frame wavelengths for each filter.

Based on these results, we calculated $\tau_{\mathrm{\lambda,irr}}$ using Equation~\ref{ssd_eq}, while $\tau_{\mathrm{\lambda,diff}}$ was taken to be the rest-frame H$\beta$ BLR lag. The total RPC model-predicted time delays, $\tau_{\mathrm{\lambda, tot}}$, were then obtained from Equation~\ref{eq:rpc} for a covering factor of 0.2 as suggested by \citet{netzer2022}. These predictions are shown in bottom panel of Figure~\ref{fig:model} as red square open points, while the observed original data are represented by black points.

Overall, the model reproduces the observed delays around the Balmer jump ($\tau_{bM2}$, $\tau_{bW1}$, $\tau_{bU}$, $\tau_{bu}$, $\tau_{bv}$, $\tau_{bB}$, $\tau_{bb}$, $\tau_{bV}$, $\tau_{by}$) reasonably well within errors, with a $\chi^2$/dof of 2.19, when assuming no free parameters. However, it fails to accurately match the lags in certain filters, most notably $\tau_{bW2}$ and $\tau_{bI}$. In particular, near the Paschen jump, the recovered lag $\tau_{bI}$ is substantially underestimated. This result suggests that the contribution from the BLR diffuse continuum is significantly underestimated when adopting a covering factor of 0.2, while the contribution from disk emission remains negligible. However, increasing the covering factor would substantially enhance the Balmer jump, which in turn would produce unrealistically large values for both $\tau_{bU}$ and $\tau_{bu}$.  In addition, the model predicts a significantly larger $\tau_{bW2}$ than observed. This discrepancy can be attributed to the inclusion of an additional fraction of BLR emission line contamination in the $W2$ band (see top panel of Figure~\ref{fig:model}), which leads to an increased diffuse emission lag in the model. In contrast, the observed spectrum shows contributions from the narrow C III] 1909 {\AA} line and a fraction of the broad C IV 1550 {\AA} line within the $W2$ band (see Figure \ref{fig:spec}). Since broadband filter light curves are generally dominated by the continuum, the bias introduced by these emission line contributions in lag measurements is expected to be small \citep{2016ApJ...821...56F}. Similar discrepancies between the observed data and RPC model predictions have also been reported by \cite{edelson2024}. However, the soft X-ray excess below $\sim$1–2 keV, commonly observed in AGN X-ray SEDs, likely originates from Comptonization in a warm, optically thick ($\tau_{\text{op-depth}} > 10$), and relatively cool ($kT \sim 0.1$–1 keV) corona located on or above the disk \citep{2018A&A...611A..59P}. Photons emitted from this extended region can irradiate the outer disk, increasing the effective light-travel distance between the variable illuminating source and the disk annuli that produce UV and optical emission. Furthermore, reprocessing within the warm corona on thermal or Comptonization timescales smooths rapid variability \citep{2017MNRAS.470.3591G, kubota2018}, collectively leading to longer negative lags ($\tau_{bW2}$ and $\tau_{bM2}$) between optical and UV bands than predicted by the RPC model, which does not include contributions from the warm corona.

These results suggest that, although the RPC model reasonably reproduces the time delays around the Balmer jump, it fails to capture the full lag-spectrum of Fairall~9 with high accuracy, possibly due to its assumption of negligible or minimal contributions from the irradiated disk, along with its incomplete treatment of emission line features and the diffuse continuum near the Paschen jump.

However, the significant underestimation in the model-recovered lag $\tau_{bI}$ is likely due to the omission of emission from hot dust in the torus ($L_{\mathrm{dust}}$), which contributes predominantly at longer optical wavelengths. To account for this component, we first generated the spectral shape of dust emission at a temperature of $T_{\mathrm{dust}}=1700$ K, normalized to the mean observed ratio $L_{5100}/L_{2\mu m}$. This yielded the spectral shape of $\frac{L_{\mathrm{dust}}}{L_{\mathrm{inc}}+L_{\mathrm{diff}}}$, shown as the dashed brown line in the top panel of Figure~\ref{fig:model}. We then incorporate this dust contribution into the continuum time delay, leading to the following expression:

\begin{equation}
\label{eq:rpc2}
\begin{split}
\tau_{\lambda, \mathrm{tot}} 
  = \tau_{\lambda, \mathrm{irr}}
    \bigg(\frac{L_{\mathrm{inc}}}{L_{\mathrm{inc}}+L_{\mathrm{diff}}+L_{\mathrm{dust}}}\bigg) 
  + \tau_{\lambda, \mathrm{diff}}
    \bigg(\frac{L_{\mathrm{diff}}}{L_{\mathrm{inc}}+L_{\mathrm{diff}}+L_{\mathrm{dust}}}\bigg) \\
  + \tau_{\lambda, \mathrm{dust}}
    \bigg(\frac{L_{\mathrm{dust}}}{L_{\mathrm{inc}}+L_{\mathrm{diff}}+L_{\mathrm{dust}}}\bigg).
\end{split}
\end{equation}

Here, $\tau_{\lambda, \mathrm{dust}}$ is taken to be approximately four times the H$\beta$ BLR lag, corresponding to the $K$-band dust lag \citep{2006ApJ...639...46S, 2014ApJ...788..159K, 2020MNRAS.491.4615K, 2024ApJ...968...59M}. The resulting time delays are shown as solid red squares in the bottom panel of Figure~\ref{fig:model}, assuming a BLR covering factor of 0.2 and a dust covering factor of 0.4. With this inclusion, the lags around the Balmer jump and $\tau_{bI}$ are successfully reproduced, indicating a significant contribution from hot dust within the $I$ filter, as expected. The reduced $\chi^2$/dof of 0.67 confirms a significantly improved fit. Nevertheless, the lags $\tau_{bW2}$ and $\tau_{bM2}$ remain substantially larger than observed.

\begin{table}[h!]
\centering

 \caption{Relativistic accretion disk model fitting results}
 \label{tab:mod_rel}

\resizebox{9cm}{!}{
\fontsize{12pt}{12pt}\selectfont
\begin{tabular}{ccccr} \hline \hline

 case & spin & $h$ ($r_g$) & $\lambda_{\text{Edd}}$   
\\ 
(1) & (2) & (3) & (4)
\\ \hline \\
full lag-spectrum & $a\ast=0$& 20.0 [11.5, 35.3] & 0.080 [0.019, 0.647] \\
& $a\ast=0.998$& 20.0 [8.5, 54.1] & 0.238 [0.034, 1.025]
\\ \\
lag-spectrum without $\tau_{bI}$ & $a\ast=0$& 19.7 [10.2, 33.9] & 0.035 [0.009, 0.161] \\
& $a\ast=0.998$& 29.4 [19.1, 41.5] & 0.043 [0.019, 0.194]
\\
\hline
\end{tabular}
}

\tablefoot{Columns are: (1) dataset used for model fitting, (2) spin, (3) median corona height in unit of gravitational radius ($r_g$), and (4) median of accretion rate. Ranges of the best-fit parameters $h$ and $\lambda_{\text{Edd}}$ for different values of $L_{\mathrm{ext}}$ (0.1, 0.5, 0.9) and color corrections $f_{\mathrm{col}}$ (1, 1.7, 2.4), listed in square brackets for each spin configuration.}

\end{table}

\subsection{Relativistic accretion disk model including X-ray reflection}
In addition, we applied a relativistic accretion disk model with realistic X-ray reflection, implemented in the code {\tt KYNXiltr} \citep{kammoun2021, kammoun2023}, to the lag-spectrum data of Fairall~9. In this framework, an X-ray point-like source illuminates a standard Novikov-Thorne accretion disk \citep{1973blho.conf..343N}, as described in the lamp-post geometry. A fraction of the incident X-ray radiation is subsequently reprocessed and emitted as UV-optical continua, producing the observed inter-band time delays. We fitted the lag-spectrum of Fairall~9 using {\tt KYNXiltr}.

During the fitting procedure, we fixed $M_{\text{BH}}$ to $2.18 \times 10^8 \, M_{\odot}$ and explored three values of the parameter, $L_{\text{ext}} = 0.1, 0.5$, and $0.9$. The parameter, $L_{\text{ext}}$, parametrizes the X-ray luminosity as the ratio of the accretion power released within a given radius to the total accretion power. In addition, we considered three values of the color correction factor, $f_{\text{color}} = 1.0, 1.7$, and $2.4$. The accretion rate and corona height were left as free parameters. The median best-fit results for two spin configurations, $a\ast = 0$ and $a\ast = 0.998$, are summarized in Table~\ref{tab:mod_rel}, while the corresponding model-predicted time-delay dependencies on wavelength are shown in Figure~\ref{fig:model} by solid black and dashed blue lines, respectively. We find the median value of the corona height to be 20 $r_g$, with a range spanning from 11.5 to 35.3 $r_g$ (8.5 to 54.1 for $a\ast=0.998$). The Eddington ratio varies between 0.019 and 0.647 (0.034 and 1.025) with a median of 0.080 (0.238) for $a\ast=0$ ($a\ast=0.998$). For comparison, the observed Eddington ratio of Fairall~9, inferred from $M_{\text{BH}}$ and $L_{5100}$, is 0.028.

The results for the two spin configurations are broadly consistent with a reduced $\chi^2$/dof value of 1.25 when no free parameters are assumed, and 1.53 when two free parameters (i.e., the accretion rate and corona height) are included. In both cases, the recovered lags follow a power-law dependence on wavelength, consistent with expectations from the Novikov–Thorne accretion disk model, which retains the same asymptotic temperature$-$radius scaling as the standard thin disk but incorporates general relativistic corrections. However, the model fails to reproduce the observed lags near the Balmer jump and significantly underestimates the lag $\tau_{bI}$ around the Paschen jump.

 To evaluate the impact of this discrepancy, we refitted the data using the same input parameters but excluded $\tau_{bI}$ from the lag-spectrum, as it deviates significantly at the longest wavelength. The resulting best fits, shown by the dotted lines for the two spin configurations, exhibit a relatively flatter lag-wavelength dependence than before but still fail to reproduce the lags around the Balmer jump, resulting in reduced $\chi^2$/dof values of 1.87 and 2.06 for $a\ast=0$ and $a\ast=0.998$, respectively, assuming two free parameters. Nevertheless, the recovered Eddington ratios 0.035 for $a\ast=0$ and 0.043 for $a\ast=0.998$ (see Table~\ref{tab:mod_rel}) are found to be more consistent with the observed value of 0.028.

Neither the RPC model nor the relativistic accretion disk model is able to reproduce the observed lag-spectrum of Fairall~9 accurately. Though the RPC model with additional dust emission successfully captures the atomic features in the lag-spectrum and provides a better fit to the delays at longer wavelengths (as indicated by the improved reduced $\chi^2$/dof), it substantially overestimates the delays at the shortest wavelengths, namely $\tau_{bW2}$ and $\tau_{bM2}$. Furthermore, it is important to note that $\tau_{\lambda, \mathrm{irr}}$ in Equation~\ref{eq:rpc2} carries additional uncertainty due to the $\chi$ factor, which can significantly affect the recovered time delays. The relativistic accretion disk model also falls short, as the intrinsic lag-spectrum does not follow a simple power-law behavior; instead, there is strong evidence of contamination from BLR scattering and reprocessing, which this model entirely neglects.

\section{Summary}
\label{ss:summ}

We present the first photometric monitoring results  carried out on Fairall~9  from our AGN monitoring program, HALO, aimed at constraining the Hubble constant through continuum time-delay fitting. Fairall~9 was monitored over a period of 146 days with a typical cadence of 0.3 days in the $u$, $v$, $b$, $y$, and $I$ filters. Using these data, we measured inter-band time delays employing several independent methods and subsequently ICCF lag-measurements were used to construct the lag–spectrum of Fairall~9. This analysis also incorporates additional re-analyzed measurements from the literature which significantly increases the wavelength coverage in the lag-spectrum. The main findings obtained from the current study are summarized below.

\begin{enumerate}

    \item Fairall~9 exhibited significant flux variability across all filters during our observing period. To separate the nuclear and host contributions, we employed the FVG method, which enabled us to estimate the host-galaxy flux and thereby recover the intrinsic AGN flux. The FVG analysis further indicates that the host galaxy of Fairall~9 closely resembles an Sb-type morphology, suggesting that an intermediate-type spiral stellar population is the dominant contributor to the host emission.
    
    \item The lag–spectrum of Fairall~9 reveals significant excess lags in the $u$ and $U$ bands, clearly indicating a contribution from the diffuse Balmer continuum emission originating in the BLR.
    
    \item In addition, the $I$ band exhibits noticeably larger delays in the lag-spectrum, which can be attributed to a potential contribution from dust emission in the torus as well as from the Paschen jump originating in the BLR.
    
    \item A conventional lag–spectrum fit, assuming a simple power-law dependence of continuum delay on wavelength, departs significantly from the predictions of the SSD model. Both the power-law slope ($\beta$) and the disk size inferred from the fit are larger than expected for Fairall~9. When the most deviant $I$-band lag is excluded from the lag-spectrum, however, the fitted slope decreases significantly to $\beta = 0.50$, except in the original-total case where a steeper slope of $\beta = 1.83$ is obtained. Nevertheless, because of the Balmer jump and the contamination in the measured fluxes by scattering and reprocessing within the BLR, such a simple power-law fit does not fully capture the features of the lag-spectrum and is therefore inadequate for detailed analysis, particularly for the observed lag-spectrum of Fairall~9.
    
    \item  To explore more physically motivated scenarios, we applied two existing models to the lag–spectrum of Fairall~9: (i) the RPC model, which accounts for a dominant contribution from diffuse BLR emission to the inter-band continuum delays, and (ii) a relativistic accretion disk model, which incorporates realistic X-ray reflection to explain the observed UV–optical time delays. We find that the RPC model reproduces the observed delays reasonably well across most wavelengths, except in the shortest-wavelength bands, $W2$ and $M2$, provided that hot dust emission is included in addition to the disk and BLR contributions.  In contrast, the relativistic accretion disk model fails to accurately reproduce the inter-band lags. Nevertheless, both models are sensitive to the assumed source luminosity and therefore cannot be used to constrain $H_0$, which is the ultimate goal of this dataset to be explored in future work.

\end{enumerate}

%\section{Data availability}
 
%The data presented in Table~\ref{tab:lc} are available only in electronic form at the CDS via \url{http://cdsweb.u-strasbg.fr/cgi-bin/qcat?J/A+A/}.

\begin{acknowledgements}
We thank the referee for comments and suggestions. This project has received funding from the European Research Council (ERC) under the European Union’s Horizon 2020 research and innovation program (grant agreement No. [951549]). The Czech-Polish Mobility program of the two Academies of Sciences, titled
“Appearance and dynamics of accretion onto black holes”, is greatly appreciated. VKJ acknowledges the OPUS-LAP/GA ˇCR-LA bilateral project (2021/43/I/ST9/01352/OPUS
22 and GF23-04053L). MHN also acknowledges the financial support by the University of Liege under Special Funds for Research, IPD-STEMA Program. SP is supported by the international Gemini Observatory, a program of NSF NOIRLab, which is managed by the Association of Universities for Research in Astronomy (AURA) under a cooperative agreement with the U.S. National Science Foundation, on behalf of the Gemini partnership of Argentina, Brazil, Canada, Chile, the Republic of Korea, and the United States of America. BC and SP acknowledge the support
from COST Action CA21136 - Addressing observational tensions in cosmology with systematics and fundamental physics (CosmoVerse), supported by COST (European Cooperation in Science and Technology). FPN gratefully acknowledges the generous and invaluable support of the Klaus Tschira Foundation. We also acknowledge support from the Polish Ministry of Science and Higher Education grant 2024/WK/02. MZ received the support from the Czech Science Foundation Junior Star grant no. GM24-10599M.
\end{acknowledgements}

\bibliographystyle{aa}
\bibliography{main}

\begin{appendix}
\label{ss:apendix}
\onecolumn

\section{Flux variation gradient}
\label{ap:fvg}

\begin{figure*}[h!]
    \centering
    \includegraphics[width=\columnwidth]{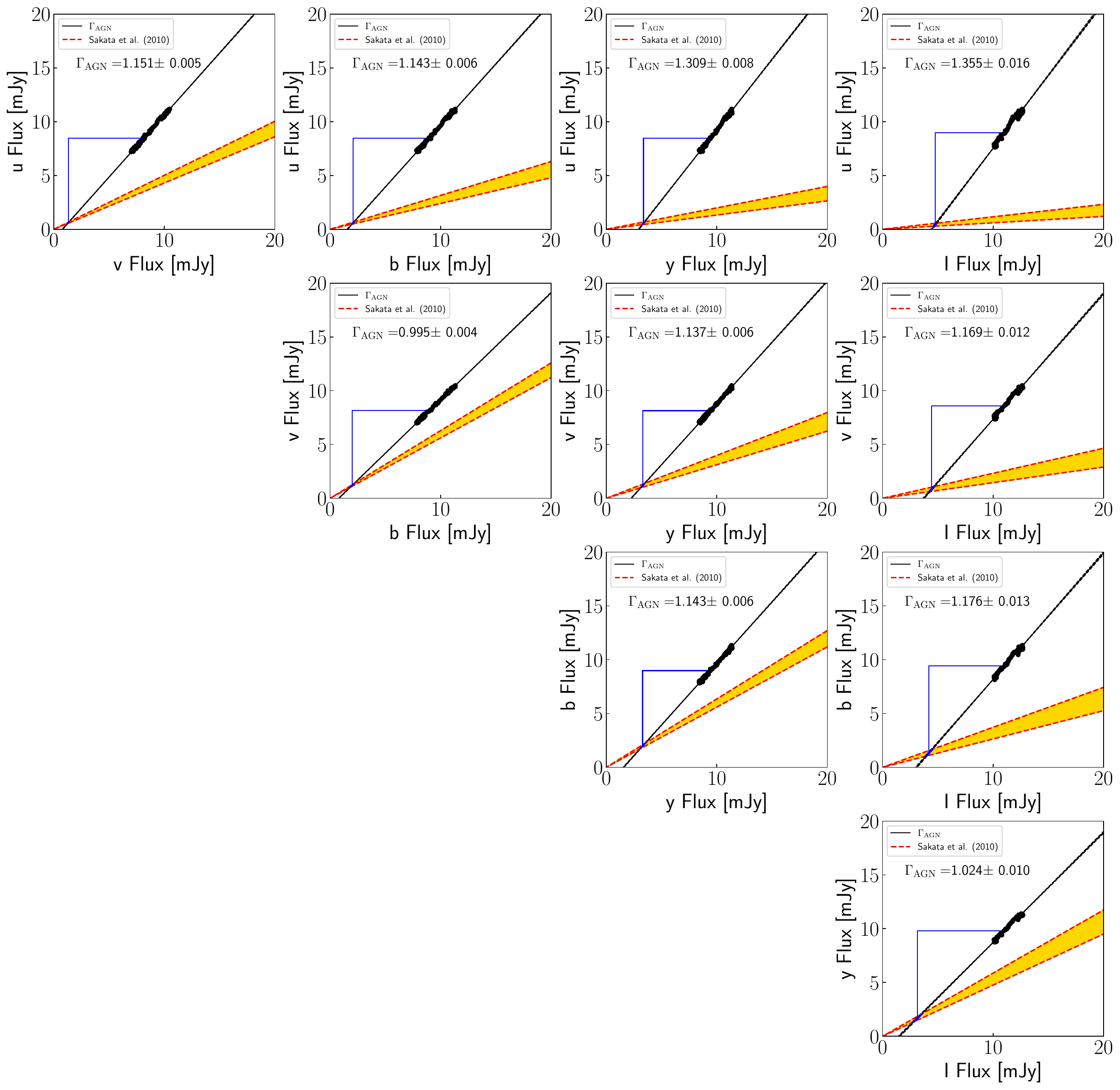}\
    \caption{Flux variation gradient (FVG) diagrams for all filter pairs. 
Black points are the observed fluxes; the OLS-bisector fit (black line) gives the AGN slope $\Gamma_{\rm AGN}$ (values in each panel). 
Dashed red lines show the range of host slopes from \citet{2010ApJ...711..461S} (yellow region); their intersections with $\Gamma_{\rm AGN}$ yield the host fluxes. 
Blue lines indicate the corresponding AGN flux ranges.}
    \label{fig:fvg}
\end{figure*}

Fig.~\ref{fig:fvg} shows the FVG diagrams for all filter pairs used in the decomposition discussed in Sect.~\ref{host-sub}. The points follow tight linear relations, indicating variability dominated by a single continuum component. The OLS-bisector slopes, $\Gamma_{\rm AGN}$ (reported in each panel), are consistently $>1$, implying a bluer variable AGN continuum relative to the host. The dashed red lines mark the empirical range of host-galaxy slopes from \citet{2010ApJ...711..461S} (yellow locus). Intersections between $\Gamma_{\rm AGN}$ and the host range yield the host flux in each band, while the corresponding AGN-only ranges are traced by the blue lines. The small scatter about the regressions indicates negligible contamination from emission lines or variable extinction within our uncertainties.

\section{Effect of detrending}
\label{ss:eff_detrend}

In this section, we investigated the potential impact of detrending AGN photometric continuum light curves to remove long-timescale variations arising either from stochastic, multi-scale accretion-disk variability or from reprocessing in the BLR and torus. To this end, we incorporated previously published light curves of Fairall~9 from \citet{edelson2024}, who provided multi-band monitoring in Swift-$HX$, $SX$, $W2$, $M1$, $W1$, $U$, $B$, and $V$ with a typical cadence of $\sim$1.3 days over a baseline of $\sim$1.8 years. Among these, the $U$-band light curve spans a longer baseline of $\sim$2.7 years, from 13 May 2018 to 14 February 2021, and overlaps with the wavelength coverage of the $u$-band used for our observations.

\begin{figure*}[h!]
    \centering
   \includegraphics[width=0.8\linewidth]{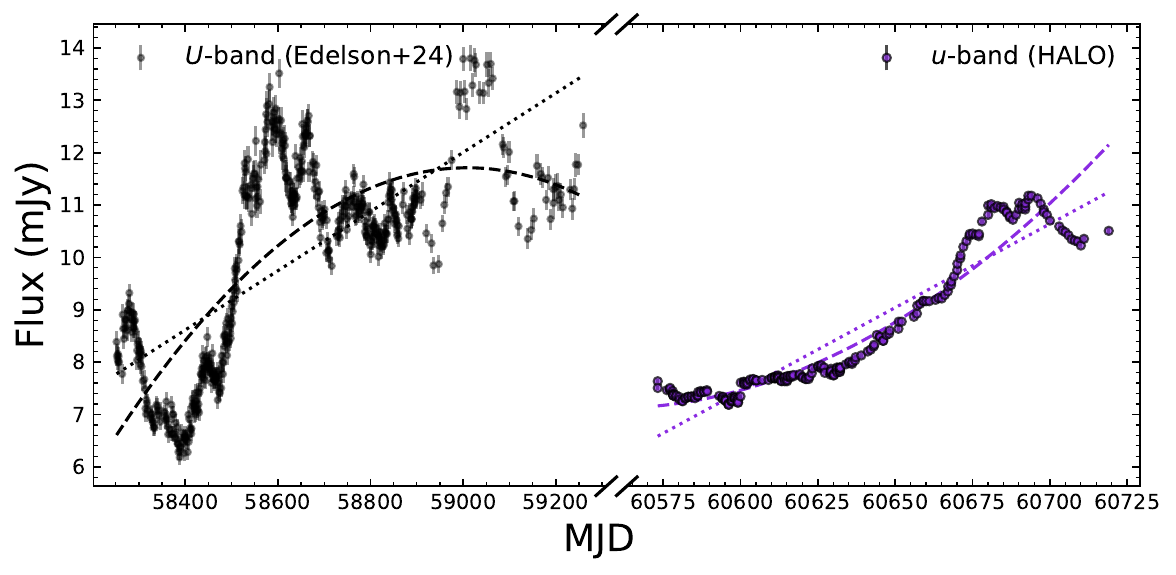}
    \caption{Comparison of the light curves of Fairall~9 in the Swift-$U$ band (black points) from \citet{edelson2024} (MJD 58251.6570$-$59259.1066) and the OCM-$u$ band (blue-violet points) obtained during our HALO observing campaign (MJD 60573.2523$-$60719.0495), together with their respective detrending trends. The dotted and dashed lines show the detrending trends with order= 1 and 2, respectively. The Swift-$U$ and OCM-$u$ light curves exhibit distinctly different long-term detrending behaviors across their independent monitoring periods. This suggests that the detrending trend is strongly dependent on the overall duration of the light curve, as different time spans subtract different components of the slower variability from the observed signal.}
    \label{fig:detrend_lc}
\end{figure*}

Our monitoring campaign covers the period from 20 September 2024 to 13 February 2025, with a duration of $\sim$146 days and a much denser cadence of $\sim$0.3 days. In Figure~\ref{fig:detrend_lc}, we present the $U$-band light curve from \citet{edelson2024} and our $u$-band light curve. We then applied first- and second-order detrending procedures to each light curve independently. Interestingly, the two datasets, which span distinct and non-overlapping monitoring periods, exhibit different detrending behaviors; in particular, a completely different trend is observed under second-order detrending.

This contrast demonstrates that the shape of the detrending curve is highly sensitive to the total length of the light curve. Consequently, detrending may subtract different components of the long-timescale, slower variability depending on the duration of the dataset. This, in turn, demonstrates that detrending the light curve can compromise the reliable recovery of the intrinsic short-timescale accretion-disk reverberation signal. Therefore, we adopt the lag analysis results obtained from the undetrended light curves as our final measurements, while those derived from detrended light curves are presented for comparison.

\section{Lag analysis results for light curves from HALO observations}
\label{ap:lag}
We employed the ICCF, {\tt PyROA}, and {\tt JAVELIN} methods to estimate the time lags between the $u$-band light curve and those obtained in the $v$, $b$, $y$, and $I$ bands from the OCM observations conducted as part of the HALO program. Lag measurements were performed using multiple data combinations, both with and without the application of detrending, following the procedures described in Sections~\ref{ss:detrend} and~\ref{lag_analysis}. The resulting lag estimates from these analyses are given in Table~\ref{tab:lag}.

\begin{table*}[h!]
\centering

 \caption{Results of lag analysis for the entire data (Total), for part 1 (S1), and for part 2 (S2)}
 \label{tab:lag}

\resizebox{19cm}{!}{
\fontsize{30pt}{30pt}\selectfont
\begin{tabular}{ccccccc} \hline \hline

Data & method & $\tau_{uu}$ & $\tau_{uv}$ & $\tau_{ub}$ & $\tau_{uy}$  & $\tau_{uI}$ 
\\ 
 &  & (days) &  (days) & (days) & (days) & (days) \\
(1) & (2) & (3) & (4) & (5) & (6) & (7)
\\ \hline \\

&  &  & Total &  &  &  \\ \\

Original & ICCF & $-0.00_{-0.92}^{+0.96}$ (1.00)  & $-0.22_{-1.08}^{+0.97}$ (1.00) & $-0.86_{-0.93}^{+0.95}$ (1.00) & $0.42_{-0.75}^{+1.27}$ (1.00) & $7.37_{-1.50}^{+1.14}$ (0.99) \\
Original & {\tt PyROA} & $0.00_{-0.05}^{+0.05}$  & $0.39_{-0.09}^{+0.09}$ & $0.11_{-0.09}^{+0.10}$  & $0.13_{-0.13}^{+0.13}$ & $3.68_{-0.24}^{+0.25}$ \\
Original & {\tt JAVELIN} & $-0.01_{-1.63}^{+1.59}$  & $1.32_{-0.67}^{+0.15}$ & $0.48_{-0.13}^{+0.11}$  & $1.61_{-1.10}^{+3.50}$ & $6.49_{-0.74}^{+0.89}$ \\ \\

Detrending (order=1) & ICCF & $0.00_{-0.26}^{+0.44}$ (1.00) & $0.75_{-0.26}^{+0.46}$ (0.98) & $0.75_{-0.27}^{+0.45}$ (0.98) & $0.50_{-0.45}^{+0.46}$ (0.97) & $4.02_{-0.75}^{+0.69}$ (0.82) \\

Detrending (order=1) & {\tt PyROA} & $0.00_{-0.05}^{+0.04}$  & $0.13_{-0.08}^{+0.09}$ & $0.02_{-0.04}^{+0.06}$  & $-0.03_{-0.07}^{+0.05}$ &  $5.08_{-0.27}^{+0.27}$  \\ 

Detrending (order=1) & {\tt JAVELIN} & $-0.00_{-2.51}^{+2.45}$  & $1.25_{-0.31}^{+0.17}$ & $0.51_{-0.12}^{+0.15}$  & $1.08_{-0.53}^{+0.62}$ &  $5.45_{-0.86}^{+0.08}$  \\ \\

Detrending (order=2) & ICCF & $0.00_{-0.45}^{+0.43}$ (1.00) & $1.21_{-0.47}^{+0.46}$ (0.98) & $1.42_{-0.46}^{+0.28}$ (0.98) & $1.00_{-0.48}^{+0.46}$ (0.96) & $4.82_{-0.67}^{+0.38}$ (0.81) \\

Detrending (order=2) & {\tt PyROA} & $-0.00_{-0.05}^{+0.05}$ & $0.63_{-0.09}^{+0.09}$  & $0.65_{-0.09}^{+0.09}$ & $0.44_{-0.12}^{+0.13}$ & $5.12_{-0.20}^{+0.20}$ \\

Detrending (order=2) & {\tt JAVELIN} & $0.03_{-2.74}^{+2.95}$ & $1.34_{-0.12}^{+0.16}$  & $0.74_{-0.20}^{+0.58}$ & $1.45_{-0.43}^{+3.50}$ & $6.42_{-0.91}^{+0.19}$ \\ \\

\hline
\\

&  &  & S1 &  &  &  \\ \\

Original & ICCF & $-0.01_{-1.22}^{+1.20}$ (1.00)  & $0.02_{-1.73}^{+1.86}$ (0.98) & $0.15_{-1.56}^{+1.54}$ (0.98) & $0.63_{-1.89}^{+1.11}$ (0.97) & $2.92_{-0.93}^{+0.86}$ (0.89) \\

Original & {\tt PyROA} & $-0.00_{-0.06}^{+0.06}$  & $0.26_{-0.14}^{+0.14}$ & $-0.45_{-0.12}^{+0.12}$  & $-0.34_{-0.16}^{+0.16}$ & $2.54_{-0.71}^{+0.55}$  \\

Original & {\tt JAVELIN} & $0.04_{-3.38}^{+4.29}$  & $1.28_{-0.53}^{+0.16}$ & $-1.20_{-0.27}^{+0.51}$  & $-0.76_{-0.41}^{+0.34}$ & $4.82_{-1.31}^{+2.64}$ \\ \\

Detrending (order=1) & ICCF & $0.09_{-0.62}^{+0.91}$ (1.00) & $-0.00_{-0.70}^{+0.70}$ (1.00) & $0.50_{-1.0}^{+0.77}$ (0.98) & $0.43_{-0.94}^{+0.87}$ (0.97) & $4.60_{-2.75}^{+1.20}$ (0.95) \\ 

Detrending (order=1) & {\tt PyROA} & $0.00_{-0.06}^{+0.06}$  & $-0.00_{-0.06}^{+0.06}$ & $-0.35_{-0.14}^{+0.14}$  & $-0.16_{-0.17}^{+0.17}$ & $9.63_{-0.30}^{+0.23}$  \\

Detrending (order=1) & {\tt JAVELIN} & $-0.02_{-4.69}^{+3.84}$  & $-0.04_{-4.66}^{+3.22}$ & $-1.02_{-0.34}^{+0.50}$  & $-0.53_{-0.44}^{+1.47}$ &  $5.29_{-0.46}^{+1.20}$  \\  \\

Detrending (order=2) & ICCF & $-0.00_{-0.39}^{+0.39}$ (1.00) & $0.06_{-0.38}^{+0.51}$ (0.91) & $-0.00_{-0.40}^{+0.40}$ (0.92) & $-0.01_{-0.56}^{+0.54}$ (0.88) & $3.41_{-0.95}^{+2.00}$ (0.83) \\

Detrending (order=2) & {\tt PyROA} & $-0.01_{-0.06}^{+0.06}$ & $0.41_{-0.14}^{+0.14}$  & $-0.34_{-0.13}^{+0.12}$ & $-0.13_{-0.17}^{+0.18}$ & $3.79_{-0.27}^{+0.34}$ \\

Detrending (order=2) & {\tt JAVELIN} & $0.01_{-4.22}^{+4.79}$  & $1.26_{-0.31}^{+0.18}$ & $-1.12_{-0.37}^{+0.50}$  & $-0.53_{-0.53}^{+1.89}$ &  $5.18_{-0.47}^{+1.05}$  \\

\\
\hline
\\

&  &  & S2 &  &  &  \\ \\

Original & ICCF & $0.01_{-0.48}^{+0.48}$ (1.00)  & $0.99_{-0.49}^{+0.49}$ (0.99) & $0.96_{-0.63}^{+0.48}$ (0.99) & $1.25_{-0.50}^{+0.51}$ (0.99) & $3.27_{-0.70}^{+0.53}$ (0.97) \\

Original & {\tt PyROA} & $-0.00_{-0.09}^{+0.09}$  & $0.71_{-0.14}^{+0.14}$ & $0.80_{-0.13}^{+0.14}$  & $1.06_{-0.17}^{+0.18}$ & $4.24_{-0.35}^{+0.38}$  \\

Original & {\tt JAVELIN} & $-0.07_{-2.27}^{+1.71}$  & $6.14_{-3.15}^{+2.03}$ & $1.39_{-0.48}^{+0.49}$  & $6.01_{-0.80}^{+0.86}$ & $6.65_{-0.47}^{+0.53}$ \\ \\

Detrending (order=1) & ICCF & $0.01_{-0.25}^{+0.45}$ (1.00) & $0.86_{-0.37}^{+0.36}$ (0.99) & $0.77_{-0.28}^{+0.45}$ (0.98) & $0.98_{-0.48}^{+0.47}$ (0.97) & $2.44_{-0.66}^{+0.53}$ (0.93) \\ 

Detrending (order=1) & {\tt PyROA} & $-0.00_{-0.07}^{+0.07}$  & $0.78_{-0.23}^{+0.17}$ & $0.85_{-0.20}^{+0.13}$  & $0.88_{-0.22}^{+0.12}$ & $1.01_{-0.10}^{+0.14}$  \\

Detrending (order=1) & {\tt JAVELIN} & $0.02_{-1.72}^{+1.86}$  & $5.69_{-3.51}^{+2.38}$ & $1.35_{-0.48}^{+0.60}$  & $5.68_{-1.56}^{+1.06}$ &  $6.22_{-0.47}^{+0.57}$  \\ \\

Detrending (order=2) & ICCF & $-0.00_{-0.56}^{+0.57}$ (1.00) & $1.19_{-0.43}^{+0.56}$ (0.99) & $1.38_{-0.42}^{+0.55}$ (0.99) & $1.14_{-0.56}^{+0.44}$ (0.98) & $4.37_{-0.70}^{+0.58}$ (0.96) \\

Detrending (order=2) & {\tt PyROA} & $0.01_{-0.13}^{+0.14}$ & $0.99_{-0.12}^{+0.12}$  & $0.99_{-0.10}^{+0.11}$ & $0.99_{-0.11}^{+0.11}$ & $1.02_{-0.09}^{+0.14}$ \\

Detrending (order=2) & {\tt JAVELIN} & $0.05_{-1.72}^{+2.06}$  & $6.98_{-3.77}^{+2.26}$ & $1.57_{-0.43}^{+0.51}$  & $5.91_{-1.52}^{+0.99}$ &  $6.77_{-0.51}^{+0.52}$ 

\\ \\
\hline
\end{tabular}
}

\tablefoot{Columns are: (1) light curve data type, (2) lag analysis method, (3) observed-frame time lag between $u$ and $u$ bands, (4) observed-frame time lag between $u$ and $v$ bands, (5) observed-frame time lag between $u$ and $b$ bands, (6) observed-frame time lag between $u$ and $y$ bands, and (7) observed-frame time lag between $u$ and $I$ bands. The values in parentheses are the peak correlation coefficients from the ICCF.}

\end{table*}

\section{Comparison of lags from different methods}
\label{ss:lag_compare}

\begin{figure*}[h!]
\centering
\includegraphics[width=1.0\textwidth]{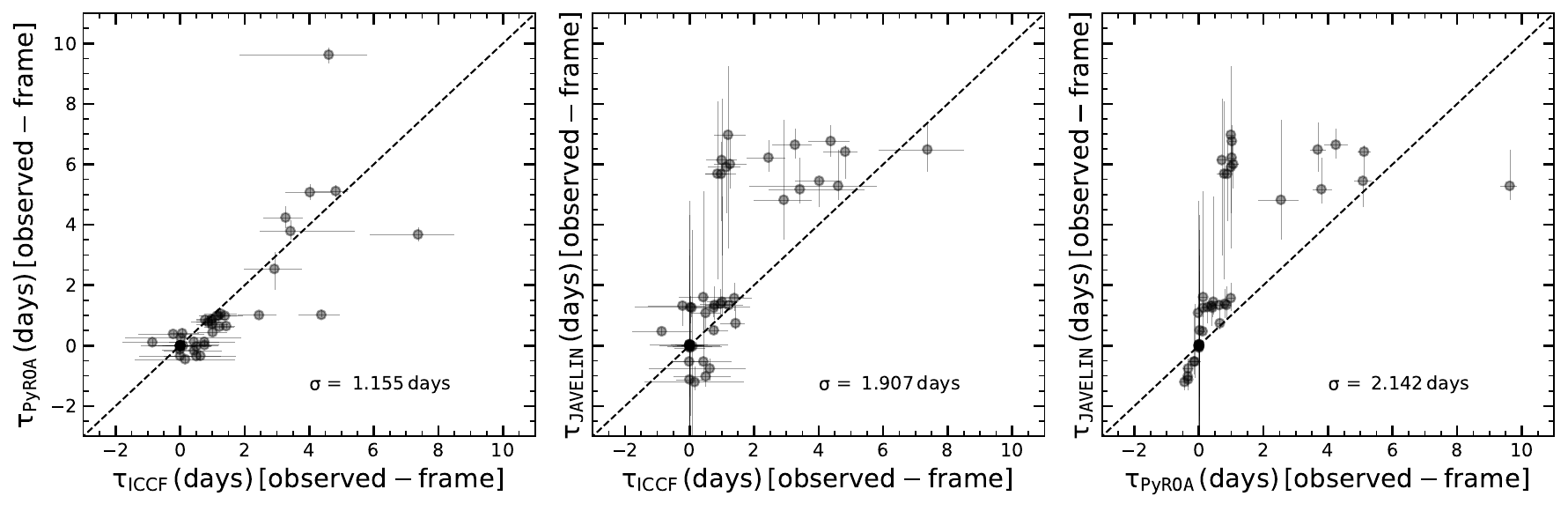}
\caption{Comparison of lag estimates between ICCF and {\tt PyROA}, ICCF and {\tt JAVELIN}, and {\tt PyROA} and {\tt JAVELIN}, in the left, middle, and right panel, respectively. The black dashed line represents the 1:1 relation. The scatter around the 1:1 line is indicated in each panel.}
\label{lag_comp}
\end{figure*}

In this section, we compared the estimated lags $\tau_{\mathrm{ICCF}}$, $\tau_{\mathrm{\tt PyROA}}$, and $\tau_{\mathrm{\tt JAVELIN}}$ obtained using three different lag analysis methods employed in this work, such as ICCF, {\tt PyROA}, and {\tt JAVELIN}, respectively. These lags were derived for various combinations of light curve data, as described in Section~\ref{lag_analysis}. The comparison included all estimated lags, as shown in Figure~\ref{lag_comp}.

To quantify the agreement between different methods, we computed the scatter around the 1:1 relation. We find a scatter of 1.155 days between $\tau_{\mathrm{ICCF}}$ and $\tau_{\mathrm{\tt PyROA}}$, 1.907 days between $\tau_{\mathrm{ICCF}}$ and $\tau_{\mathrm{\tt JAVELIN}}$, and 2.142 days between $\tau_{\mathrm{\tt PyROA}}$ and $\tau_{\mathrm{\tt JAVELIN}}$. These values indicate that $\tau_{\mathrm{ICCF}}$ and $\tau_{\mathrm{\tt PyROA}}$ show the closest agreement, while the largest discrepancy occurs between $\tau_{\mathrm{\tt PyROA}}$ and $\tau_{\mathrm{\tt JAVELIN}}$. Overall, this suggests that $\tau_{\mathrm{ICCF}}$ and $\tau_{\mathrm{\tt PyROA}}$ are more consistent with each other than either is with $\tau_{\mathrm{\tt JAVELIN}}$.

This trend is consistent with findings from \citet{2024ApJS..275...13W}, who also reported that ICCF and {\tt PyROA} typically yield more reliable lag estimates compared to {\tt JAVELIN}. However, it is important to note that {\tt PyROA} tends to underperform in modeling short-timescale rapid variations in the light curves. As discussed in Section~\ref{lag_analysis}, this limitation can lead to less reliable lag estimates than those derived from ICCF. Considering these factors, we adopt the lags estimated from ICCF as the most reliable for our analysis.

\section{Lag analysis results for light curves from Literature}
\label{lit:lag}

 It is important to note that \citet{edelson2024} performed a lag analysis of the Swift photometric light curves using the ICCF method, taking the $W2$ band as the reference. For consistency with our study, we reanalyzed the inter-band delays using the ICCF method with respect to the $U$ band, which overlaps with the Str\"omgren $u$ filter adopted as the reference band in our analysis. The obtained inter-band delays from Swift observations are given in Table~\ref{tab:lag_lit}.

\begin{table*}[h!]
\centering

 \caption{Results of Literature lag analysis}
 \label{tab:lag_lit}

\resizebox{8cm}{!}{
\fontsize{8pt}{8pt}\selectfont
\begin{tabular}{cccr} \hline \hline

Band/filter & Band center & unit &$\tau$  
\\ 
 & & & (days)  \\
(1) & (2) & (3) & (4)
\\ \hline \\

$HX$ & $3.9$  & keV & $-2.76_{-2.50}^{+2.62}$ (0.64) \\
$SX$ & $0.7$  & keV & $11.23_{-13.48}^{+10.82}$ (0.38) \\
$W2$ & $2055 \pm 357$  & {\AA} & $-4.83_{-1.32}^{+1.35}$ (0.97) \\
$M2$ & $2246 \pm 199$  & {\AA} & $-2.57_{-1.66}^{+1.91}$ (0.98) \\
$W1$ & $2580 \pm 441$  & {\AA} & $-1.87_{-1.67}^{+1.70}$ (0.99) \\
$U$ & $3463 \pm 244$  & {\AA} & $-0.05_{-1.26}^{+1.26}$ (1.00) \\
$B$ & $4350 \pm 287$  & {\AA} & $-0.46_{-1.24}^{+0.90}$ (0.99) \\
$V$ & $5425 \pm 228$  & {\AA} & $0.20_{-0.16}^{+0.16}$ (0.98) \\

\\ 
\hline
\end{tabular}
}

\tablefoot{Columns are: (1) name of the Swift bands/filters, (2) band/filter center  (observed-frame for the source), (3) unit of the band/filter central values, and (4) observed-frame inter-band time-delays with respect to $U$-band from ICCF method with their 1$\sigma$ lower and upper uncertainties. The values in parentheses represent the peak correlation coefficients from the ICCF. }

\end{table*}

\end{appendix}

\end{document}